%% file: main_clean.tex
\documentclass{aa}  
\usepackage{graphicx}
\usepackage{subcaption}
\usepackage{txfonts}

\usepackage{xcolor}
\definecolor{myred}{rgb}{0.67571825, 0, 0.17578125}
\definecolor{myblue}{rgb}{0, 0.203125, 0.3828125}
\usepackage[colorlinks=true, linkcolor=myblue, filecolor=myblue, urlcolor=myblue, citecolor=myblue]{hyperref}
\usepackage{xspace}
\usepackage{soul}
\usepackage{color}

\newcommand\PT{2024~PT$_5$\xspace}
\newcommand\Kamo{Kamo`oalewa\xspace}

\newcommand\KaretaLCDate{2024 August 14\xspace}
\newcommand\KaretaSpecDate{2024 August 16\xspace}
\newcommand\BolinSpecDate{2024 September 27\xspace}

\newcommand\MarcosSpecDate{2024 September 7\xspace}
\newcommand\MarcosLCDate{2024 September 28\xspace}
\newcommand\rad{15\xspace}
\newcommand\annin{18\xspace}
\newcommand\annout{28\xspace}
\newcommand\alphamin{14.3~deg\xspace}
\newcommand\alphamax{26.8~deg\xspace}

\newcommand\Prmin{200\xspace}
\newcommand\Prmax{8161\xspace}
\newcommand\gr{$g-r=0.567\pm0.044$\xspace}
\newcommand\ri{$r-i=0.155\pm0.009$\xspace}
\newcommand\rz{$r-z=0.147\pm0.066$\xspace}
\newcommand\grval{$0.567\pm0.044$\xspace}
\newcommand\rival{$0.155\pm0.009$\xspace}
\newcommand\rzval{$0.147\pm0.066$\xspace}
\newcommand\grKaretaPS{$0.618\pm0.029$\xspace}
\newcommand\riKaretaPS{$0.205\pm0.043$\xspace}
\newcommand\rzKaretaPS{$0.064\pm0.083$\xspace}
\newcommand\grBolinPS{$0.483\pm0.043$\xspace}
\newcommand\riBolinPS{$0.286\pm0.043$\xspace}
\newcommand\rzBolinPS{$0.007\pm0.059$\xspace}
\newcommand\grKamoPS{$0.47\pm0.04$\xspace}
\newcommand\riKamoPS{$0.20\pm0.04$\xspace}
\newcommand\rzKamoPS{$0.25\pm0.07$\xspace}
\newcommand\grlunarPS{$0.48\pm0.01$\xspace}
\newcommand\rilunarPS{$0.24\pm0.04$\xspace}
\newcommand\rzlunarPS{$0.14\pm0.11$\xspace}

\newcommand\YJbyBB{$Y-J=0.557\pm0.046$\xspace}
\newcommand\JHbyBB{$J-H=0.672\pm0.078$\xspace}
\newcommand\HKsbyBB{$H-Ks=0.148\pm0.098$\xspace}

\newcommand\HVlinear{$28.06\pm0.05$\xspace}
\newcommand\HVlinearMPC{$28.05\pm0.05$\xspace}

\newcommand\bVlinear{$0.029\pm0.002$~mag~deg$^{-1}$\xspace}
\newcommand\bVlinearMPC{$0.029\pm0.001$~mag~deg$^{-1}$\xspace}
\newcommand\pVlinear{$0.26\pm0.07$\xspace}

\newcommand\HVHG{$27.72\pm0.09$\xspace}
\newcommand\HVHGMPC{$27.78\pm0.14$\xspace}

\newcommand\GV{$0.223\pm0.073$\xspace}
\newcommand\GVMPC{$0.28\pm0.08$\xspace}

\newcommand\DiamVHG{$7.4\pm1.0$\xspace}

\begin{document} 
\title{
Multi-epoch spectro-photometric characterization of the minimoon 2024 PT$_5$ 
in the visible and near-infrared
}
\author{
Jin Beniyama\inst{1,2},
Bryce T. Bolin\inst{3},
Alexey V. Sergeyev\inst{1,4},
Marco Delbo\inst{1,5},
Laura-May Abron\inst{6},
Matthew Belyakov\inst{7},
Tomohiko Sekiguchi\inst{8},
Seiko Takagi\inst{9}
}
\institute{
Université Côte d'Azur, 
Observatoire de la Côte d'Azur, CNRS, Laboratoire Lagrange, Bd de l'Observatoire, 
CS 34229, 06304 Nice Cedex 4, France \\
\email{jbeniyama@oca.eu}
\and
The Department of Earth and Planetary Science, The University of Tokyo, 7-3-1 Hongo, Bunkyo, Tokyo 113-0033, Japan  
\and
Eureka Scientific, Oakland, CA 94602, U.S.A.
\and
Institute of Astronomy, V.N. Karazin Kharkiv National University, 35 Sumska Str., Kharkiv 61022, Ukraine
\and
School of Physics and Astronomy, University of Leicester, Leicester LE1 7RH, UK
\and
Griffith Observatory, Los Angeles, CA 90027
\and
Division of Geological and Planetary Sciences, California Institute of Technology, Pasadena, CA 91125,USA
\and
Asahikawa Campus, Hokkaido University of Education, Hokumon 9, Asahikawa, Hokkaido 070-8621, Japan
\and
Department of Earth and Planetary Sciences, Faculty of Science, Hokkaido University, Kita-ku, Sapporo, Hokkaido 060-0810, Japan 
}
\date{Received 07 08, 2025}

\abstract
{
2024~PT$_5$ is a tiny ($D\leq10$~m) near-Earth asteroid (NEA) discovered in August 2024.
2024~PT$_5$ was gravitationally bound to the Earth-Moon system from September to November 2024 and classified as a minimoon.
Several quick response observations suggest the lunar ejecta origin of 2024~PT$_5$,
while rotation state and albedo, essential properties to investigate its origin, are not well constrained.
}
{
We aim to 
characterize 
the spectro-photometric properties of 
2024~PT$_5$ by ground-based observations to test its taxonomic 
classification
and origin.
}
{
We performed visible 
to near-infrared
multicolor photometry of 2024~PT$_5$ from data taken
using the TriColor CMOS Camera and Spectrograph (TriCCS) on the Seimei 3.8~m telescope 
during 2025 January 4-10. The Seimei/TriCCS observations of 2024~PT$_5$ cover phase angles 
from 14~deg to 27~deg, and were obtained in the $g$, $r$, $i$, and $z$ bands in the Pan-STARRS system. 
In addition, we analyzed $Y$, $J$, $H$, and $K$ photometry taken with 
the Multi-Object Spectrograph for Infrared Exploration (MOSFIRE) on the Keck I 10-m telescope taken on 2025 January 16-17. 
}
{
Our lightcurves show brightness variations over time periods of several tens of minutes. 
We infer that 2024~PT$_5$ is in a tumbling state and has a lightcurve amplitude of about 0.3~mag. 
Visible and near-infrared color indices of 2024~PT$_5$,
$g-r=0.567\pm0.044$,
$r-i=0.155\pm0.009$,
$r-z=0.147\pm0.066$,
$Y-J=0.557\pm0.046$,
$J-H=0.672\pm0.078$,
and
$H-Ks=0.148\pm0.098$,
indicate that 2024~PT$_5$ is an S-complex asteroid, largely consistent with previous observations.
Using the $H$-$G$ model, we derived an absolute magnitude $H_{V,HG}$ of $27.72\pm0.09$
and a slope parameter $G_V$ of $0.223\pm0.073$ in V-band.
A geometric albedo of 2024~PT$_5$ is derived to be $0.26\pm0.07$ 
from the slope of its photometric phase curve.
This albedo value is typical of the S- and Q-type NEAs. 
}
{
Using the albedo and absolute magnitude,
the 
equivalent
diameter of \PT is estimated to be $7.4\pm1.0$~m. 
The color properties of 2024~PT$_5$ derived from our observations match rock samples taken from the lunar surface, which agrees with previous studies.
}
\titlerunning{2024 PT5}
\authorrunning{Beniyama, J., et al.}
\maketitle

\section{Introduction}
The natural bodies gravitationally bound to the Earth-Moon system,
with the exception of our Moon, 
are called minimoons.
\footnote{
In this paper, we follow the definition by, for example, \cite{Kary1996, Granvik2012, Marcos2025}: 
geocentric energy is smaller than zero, and the geocentric distance is less than three Hill radii of Earth, $\sim0.03$~au.
}
Depending on whether they are bound more than or less than 1 orbital revolution,
they are classified as temporally captured orbiters (TCO) or 
temporally captured flyby (TCF), respectively \citep[e.g.,][]{Granvik2012, Granvik2013, Jedicke2018}.
Minimoons have similar heliocentric orbits with the Earth, and have low $\Delta$v.
Thus, minimoons are possible targets of spacecraft missions,
and their characterization by telescopic observations is essential \citep{Jedicke2018}.
Only four minimoons were discovered before August 2024:
1991~VG \citep{Marcos2018},
2006~RH$_{120}$ \citep{Kwiatkowski2009},
2020~CD$_3$ \citep{Bolin2020, Fedorets2020b, Marcos2020, Naidu2021},
and 2022~NX$_1$ \citep{Marcos2023}. 
Dynamical studies have shown that lunar impacts can produce objects on Earth co-orbital orbits \citep{Gladman1995,Jedicke2025}. 
Observational studies of minimoons and Earth co-orbitals in visible to near-infrared light support the possibility that some asteroids originate as lunar ejecta  by showing spectral similarity with lunar rock samples \citep[][]{Bolin2020,Sharkey2021, Bolin2025_PT5}.

\begin{figure*}
\centering
\includegraphics[width=1.0\hsize]{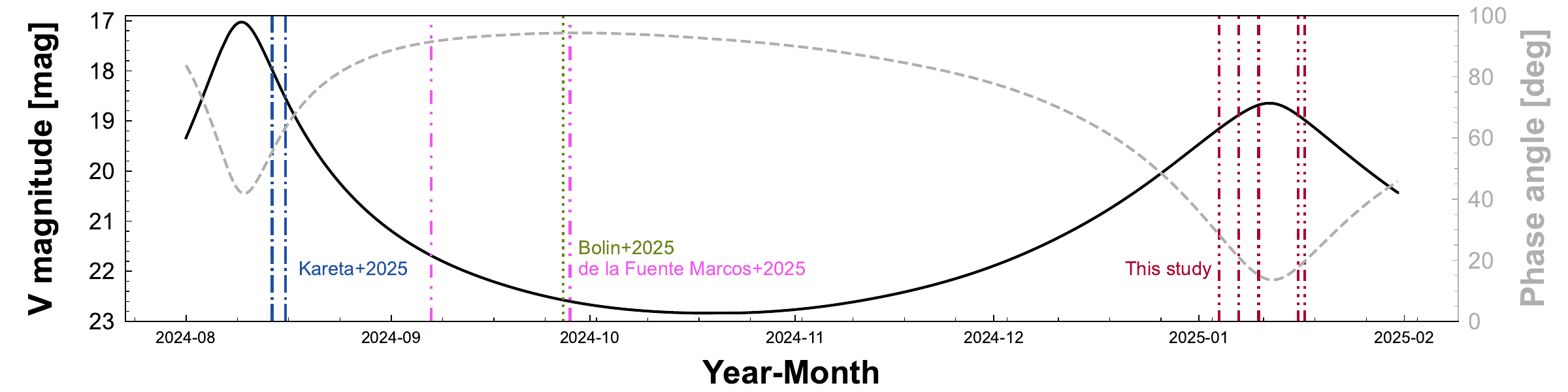}
\caption{
Predicted magnitude (solid line) and phase angles (dashed line) of \PT 
obtained from NASA JPL Horizons using the \texttt{Python} package \texttt{astroquery} \citep{Ginsburg2019}.
Vertical lines show our observations and previous observations: 
spectrophotometry and spectroscopy in \cite{Kareta2025} (dot-dashed lines),
spectroscopy and lightcurve observations in \cite{Marcos2025} (double dot–dashed lines),
spectrophotometry in \cite{Bolin2025_PT5} (dotted line),
and this study (triple dot–dashed lines).
}
\label{fig:ephem}
\end{figure*}

The target of this paper, \PT, is the fifth minimoon discovered on 2024 August 7 
by the Asteroid Terrestrial-impact Last Alert System in South Africa \citep{2024PT5MPEC,Bolin2025_PT5}.
Soon after the discovery, telescopic observations of \PT have been made worldwide 
as summarized in Table \ref{tab:existingobs}.
Fig. \ref{fig:ephem} shows the predicted magnitude and solar phase angles of \PT.

\cite{Bolin2025_PT5} performed visible to near-infrared (0.475~$\mu$m to 0.880~$\mu$m) 
\footnote{
In this paper, we follow the convention that the boundary between visible and near-infrared wavelengths is set at 0.75~$\mu$m \citep{Glass1999}.
}
spectrophotometry of \PT
using Gemini Multi-Object Spectrograph (GMOS) on Gemini North telescope on \BolinSpecDate. 
They found that the spectrum of \PT best matches lunar rock samples followed by S-complex asteroids,
making it the third known asteroid with a spectrum similar to lunar rock following 2020~CD$_3$ and (469219)~Kamo`oalewa \citep[][]{Bolin2020, Sharkey2021, Bolin2025_PT5}.

\cite{Kareta2025} 
later reported 
visible multicolor photometry of \PT using 
the Lowell Discovery Telescope (LDT) on \KaretaLCDate.
Then they activated target-of-opportunity (ToO) programs of 
visible and near-infrared spectroscopy of \PT using the LDT and NASA Infrared Telescope Facility on \KaretaSpecDate.
Their visible to near-infrared ($\sim$0.40~$\mu$m to 2.45~$\mu$m) reflectance spectrum of \PT,
covering a broad wavelength range, are in agreement with
the results of \citet[][]{Bolin2025_PT5}, showing that it is compatible with lunar samples.

Additionally, \cite{Marcos2025} characterized \PT using the 10.4~m Gran Telescopio Canarias (GTC), the Two-meter Twin Telescope (TTT),
and the Transient Survey Telescope (TST) in the Canary Islands.
Three visible to near-infrared (0.480~$\mu$m to 0.920~$\mu$m) spectra of \PT were obtained using the OSIRIS spectrograph on the GTC on \MarcosSpecDate.
Also lightcurve observations of \PT were performed using the GTC/OSIRIS on \MarcosLCDate.
Moreover, the astrometric observations were obtained
using the TTT1, TTT2, and the TST.
Their visible to near-infrared spectrum 
are also in agreement with
the results of \citet[][]{Bolin2025_PT5}, 
which showed that the visible to near-infrared (0.475~$\mu$m to 0.880~$\mu$m) spectrum of \PT is 
matches powder samples of a mare breccia of the Moon collected by the Luna 24 mission.

A rotation state of \PT is still unclear 
though the state-of-the-art ground-based telescopes were used to characterize \PT.
The rotation state is essential to investigate its origin.
For instance, the tumbling state of \PT would support the lunar ejecta origin \citep{Harris1994}. 
This is because a tumbling motion is a natural outcome after the collisional event.
\cite{Bolin2025_PT5} and \cite{Marcos2025} reported that \PT is possibly rotating with 
a rotation period shorter than $\sim$1~hr, while \cite{Kareta2025} concluded that there is no clear periodicity in their lightcurves.
Also, the geometric albedo of \PT is not estimated though it is another crucial quantity to investigate the surface property.

In this paper, we report the results of visible multicolor photometry of \PT 
for one week in 2025 January, when \PT was brighter than 20~mag in visible wavelengths 
and can be observed in a wide range of phase angles as seen in Fig. \ref{fig:ephem}
\footnote{\cite{Kareta2025} observed \PT at phase angles of $\sim64$~deg as written in their main text, not $\sim1$~deg as written in their Table 1.}. 
The paper is organized as follows.
In Sect. \ref{sec:obsred}, we describe our photometric observations and 
data reduction with Seimei/TriColor CMOS Camera and Spectrograph (TriCCS) 
and Keck/Multi-Object Spectrograph for Infrared Exploration (MOSFIRE).
The results of the observations are presented in Sect. \ref{sec:result}.
The constraints on the physical properties of \PT and its origin are discussed in Sect. \ref{sec:disc}.

\begin{table*}
\caption{\label{t7}Summary of the existing observations\label{tab:existingobs}}
\centering
\begin{tabular}{ll}
\hline\hline
Reference    & Method                                                        \\
\hline
\cite{Bolin2025_PT5}    & Visible lightcurve ($\sim$40~min)\\
& Spectrophotometry (0.475~$\mu$m to 0.880~$\mu$m)\\
\cite{Kareta2025}  & Visible lightcurve ($\sim$40~min)\\
& Spectrophotometry \& Spectroscopy ($\sim$0.40~$\mu$m to 2.45~$\mu$m)\\            
\cite{Marcos2025}  & Visible lightcurve ($\sim$1~hr)\\
& Spectroscopy (0.480~$\mu$m to 0.920~$\mu$m)\\
\hline
\end{tabular}
\end{table*}

\section{Observations and data reduction}\label{sec:obsred}
We obtained multicolor photometry of \PT in January 2025.
The observing conditions are summarized in Table \ref{tab:obs}.
The predicted $V$-band magnitudes, phase angles, distances between \PT and observer, 
and distances between \PT and the Sun in Table \ref{tab:obs} were obtained from NASA JPL
Horizons \footnote{\url{https://ssd.jpl.nasa.gov/horizons}} using the \texttt{Python} 
package \texttt{astroquery} \citep{Ginsburg2019}.

\subsection{Seimei telescope}
We observed \PT using the TriCCS 
on the 3.8~m Seimei telescope \citep{Kurita2020} on 2025 January 4, 7, and 10.
The telescope is located at the Kyoto University Okayama Observatory 
(133.5967$^\circ$ E, 34.5769$^\circ$ N, and 355~m in altitude).
We simultaneously obtained three-band images in the Pan-STARRS ($g$, $r$, $i$) 
and ($g$, $r$, $z$) filter \citep{Chambers2016}.
The field of view is $12.6\arcmin\times7.5\arcmin$ with a pixel scale of 0.350~arcsec~pixel$^{-1}$.

A phase angle of \PT changed from 28.7~deg on 2025 January 4 to 14.3~deg on 2025 January 10.
2024 PT5 had a geocentric distance of about 0.012~au and a heliocentric distance of about 0.994--0.995~au throughout  the observing period.
The lunar phases on 2025 January 4, 7, and 10 were $\sim$0.2, $\sim$0.6, and $\sim$0.9, respectively.
The lunar elongations on 2024 January 4, 7, and 10 were $\sim$120~deg, $\sim$80~deg, and $\sim$30~deg, respectively.
The apparent sky motion of \PT was about 0.13--0.15~arcsec~s$^{-1}$.
The seeing measured by using in-field stars was 2.7--3.3~arcsec in the $r$ band.

Non-sidereal tracking was performed during the observations of \PT.
Exposure times were set to 5~s for all observations.
We took multiple images with short exposures rather than a single image with 
long exposures in our observations to avoid having elongated photometric reference stars and also to eliminate the cosmic rays.
We performed standard image reduction, including bias subtraction, dark subtraction, and flat-fielding.
The astrometry of reference sources from the Gaia Data Release 2 was performed using the \texttt{astrometry.net} software \citep{Lang2010}.

\begin{figure*}[ht]
\centering
\begin{subfigure}[b]{0.24\hsize}
\includegraphics[width=\linewidth]{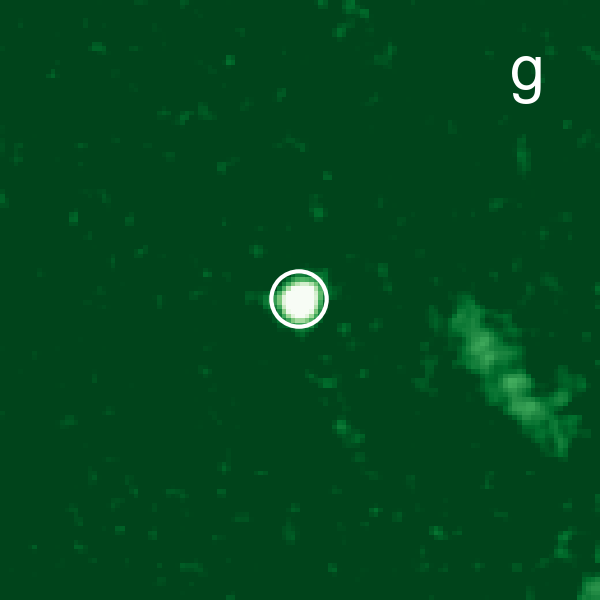}
\end{subfigure}
\begin{subfigure}[b]{0.24\hsize}
\includegraphics[width=\linewidth]{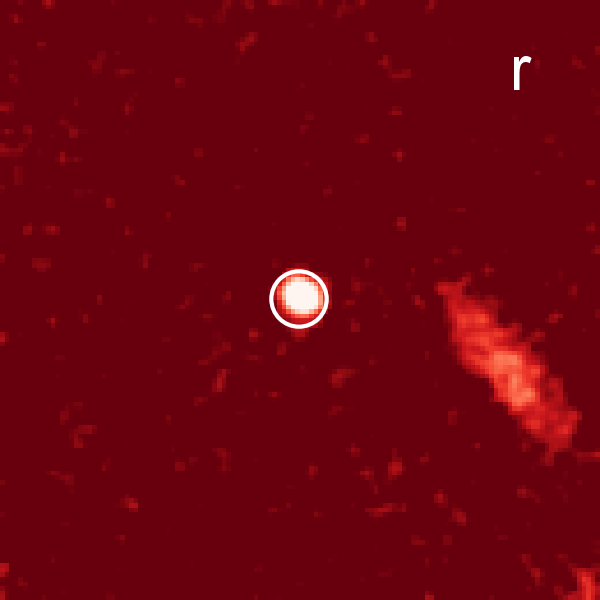}
\end{subfigure}
\begin{subfigure}[b]{0.24\hsize}
\includegraphics[width=\linewidth]{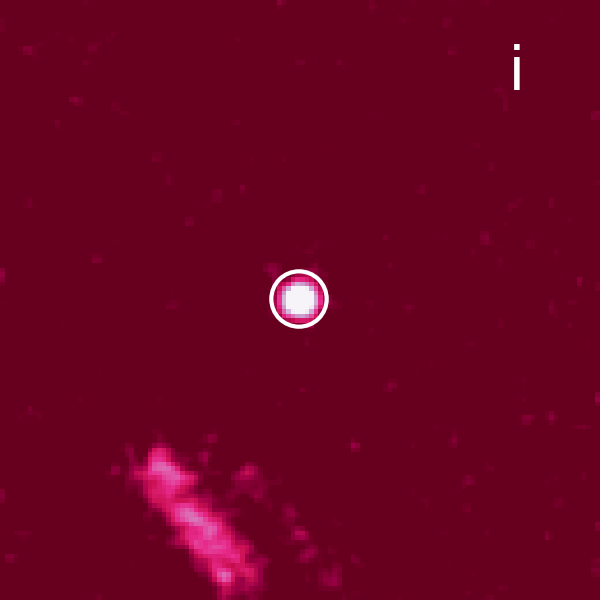}
\end{subfigure}
\begin{subfigure}[b]{0.24\hsize}
\includegraphics[width=\linewidth]{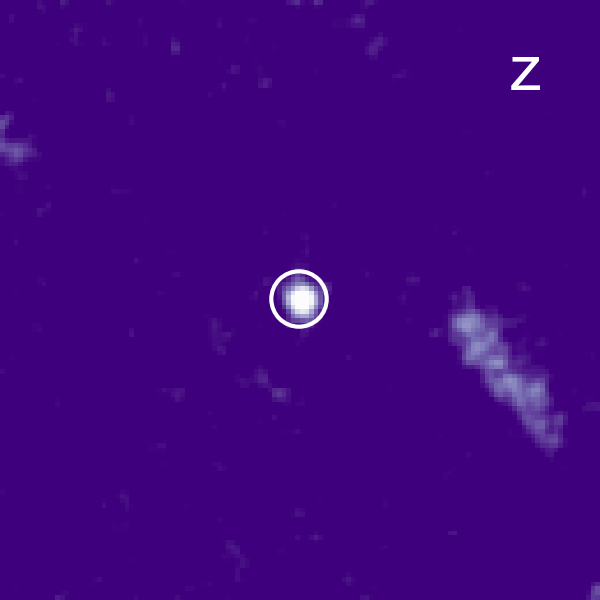}
\end{subfigure}
\caption{
Non-sidereally stacked images 
in $g$, $r$, $i$, and $z$ bands with a total integration time of 100~s on 2024 January 4 are shown. 
Circles indicate \PT.
Field of view covers 
$45~arcsec\times45~arcsec$.
North is to the top and East is to the left.
}
\label{fig:cutout_VIS}
\end{figure*}

\begin{figure*}[ht]
\centering
\begin{subfigure}[b]{0.24\hsize}
\includegraphics[width=\linewidth]{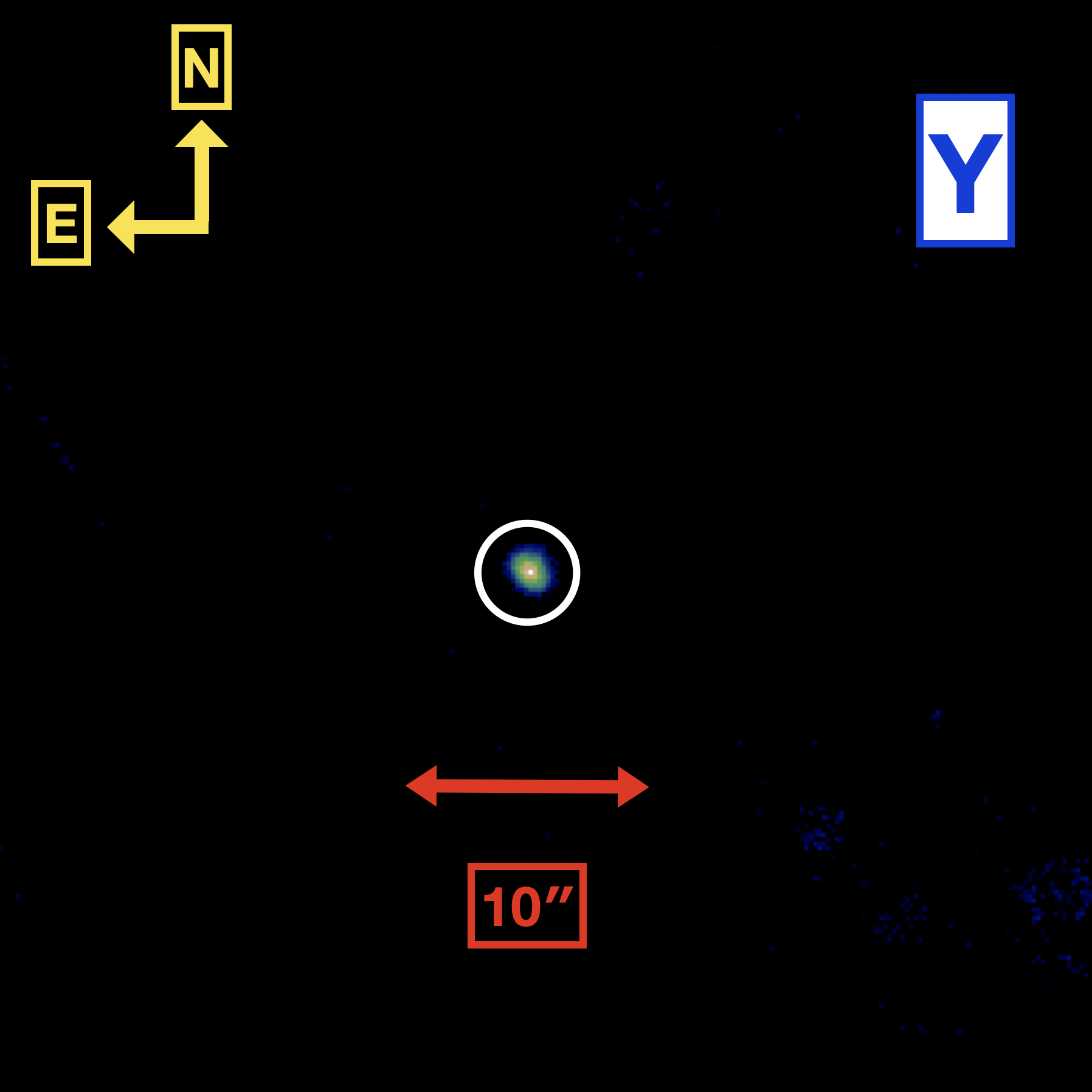}
\end{subfigure}
\begin{subfigure}[b]{0.24\hsize}
\includegraphics[width=\linewidth]{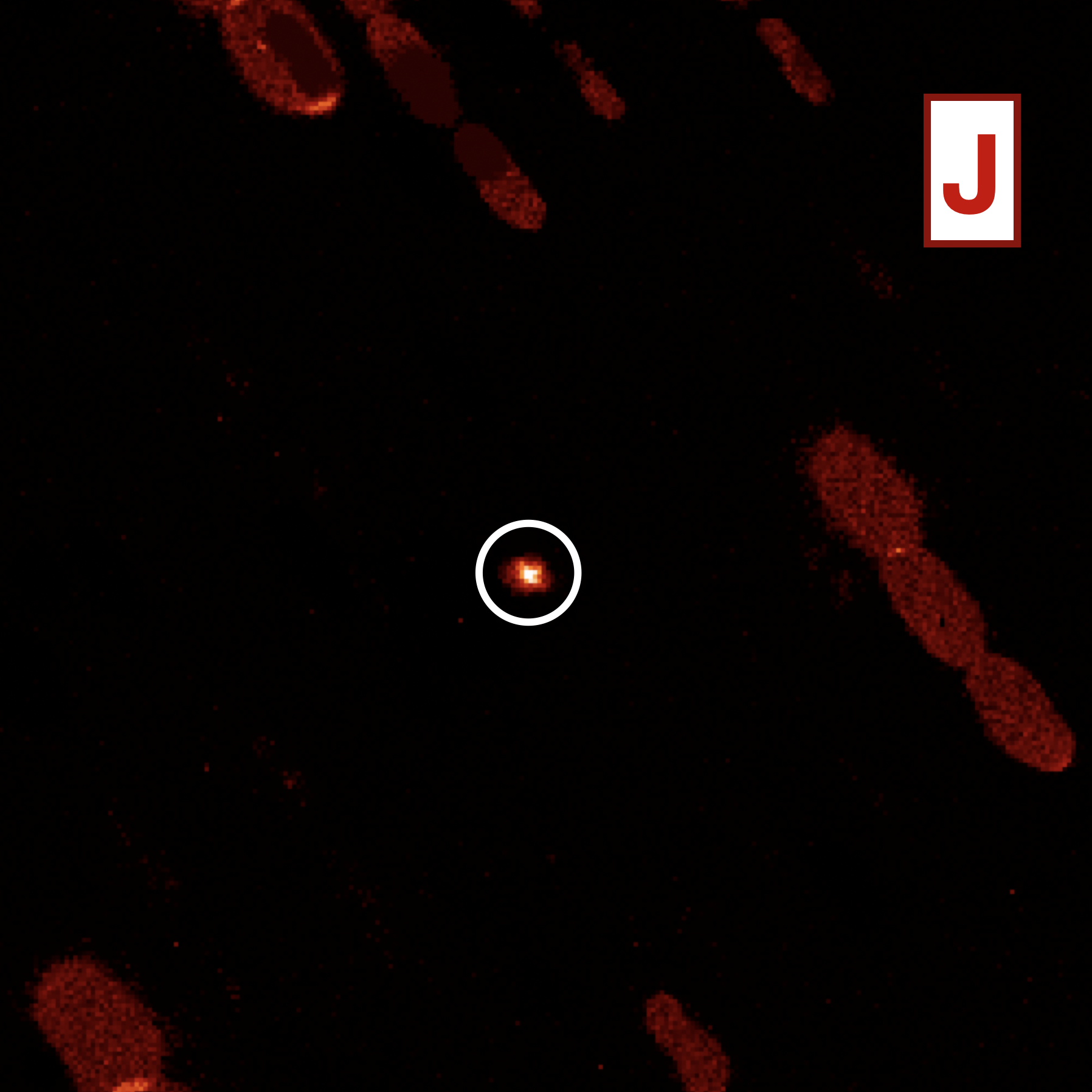}
\end{subfigure}
\begin{subfigure}[b]{0.24\hsize}
\includegraphics[width=\linewidth]{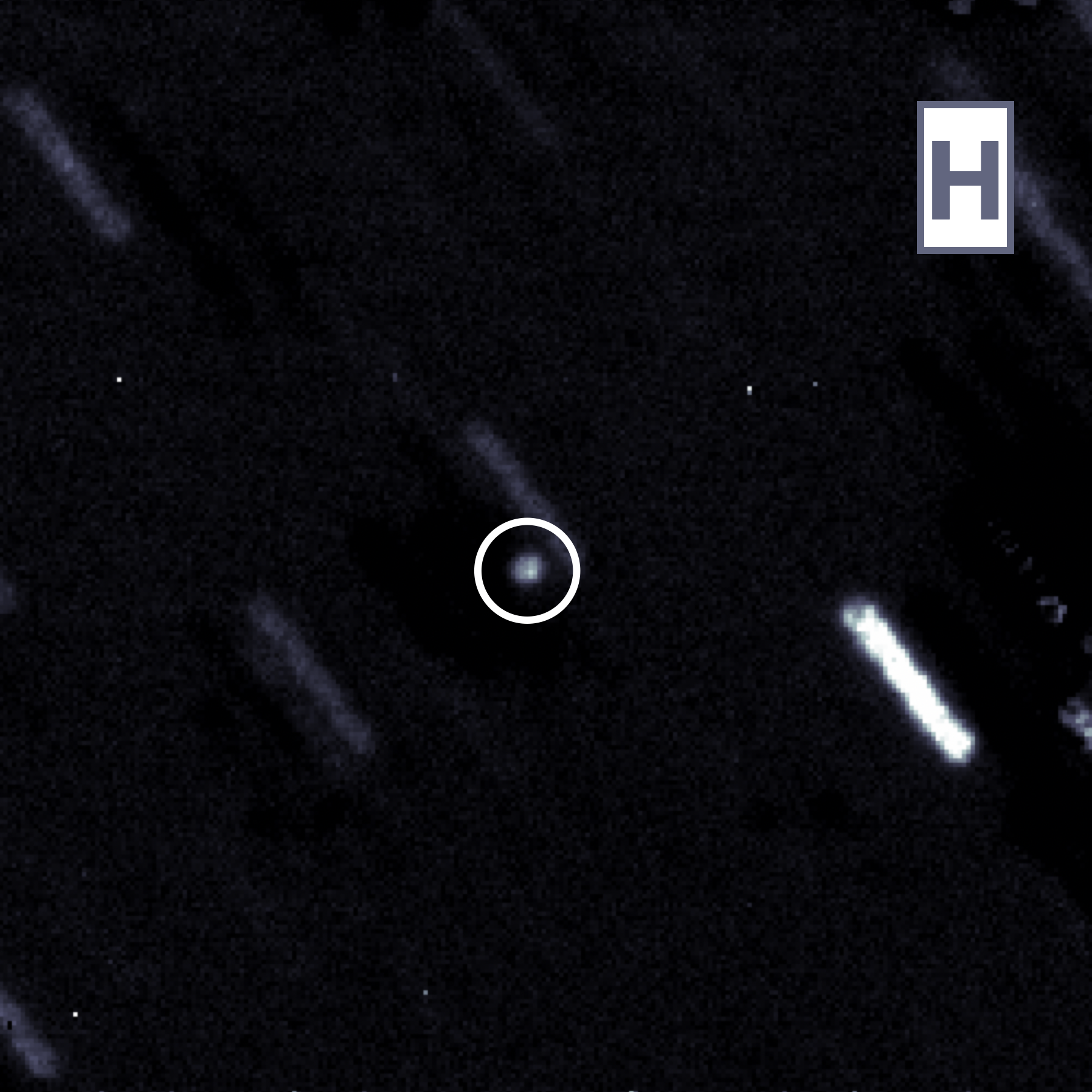}
\end{subfigure}
\begin{subfigure}[b]{0.24\hsize}
\includegraphics[width=\linewidth]{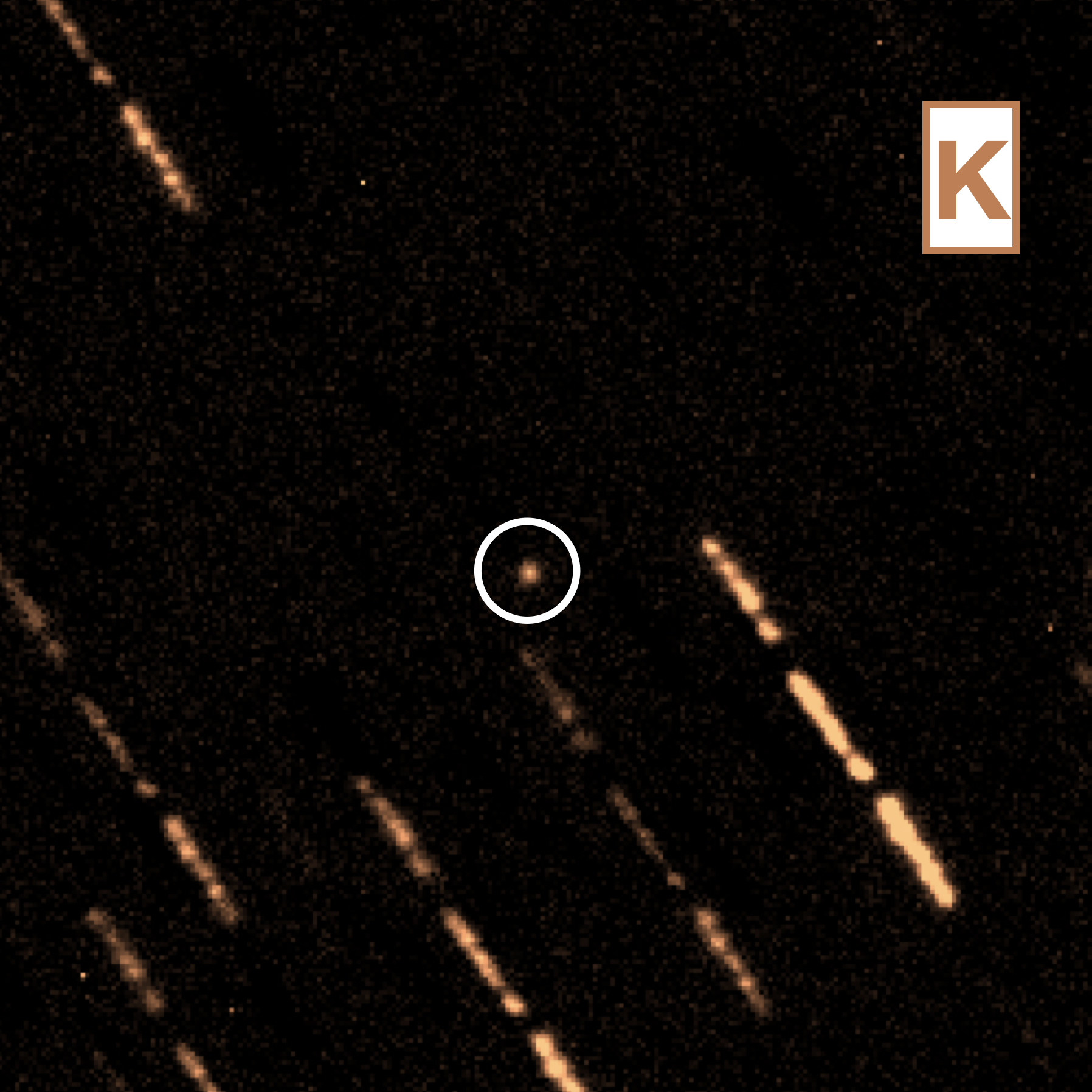}
\end{subfigure}
\caption{
Non-sidereally stacked images 
in $Y$, $J$, $H$, and $K$ bands with a total integration time of 150 s on 2024 January 16/17 are shown. 
Circles indicate \PT.
Field of view covers 
$45~arcsec\times45~arcsec$.
North is to the top and East is to the left.
}
\label{fig:cutout_NIR}
\end{figure*}

We performed stacking of images before photometry to increase the signal-to-noise ratio (S/N) 
of \PT avoiding the elongations of their images as shown in the upper panels of Fig. \ref{fig:cutout_VIS} (hereinafter referred to as the nonsidereally stacked image).
The S/N of \PT on the single 5~s images are as low as a few to less than ten.
Especially, the S/N of \PT on 2024 January 10 are lower than the others due to 
the larger lunar phase ($\sim$0.9) and smaller lunar elongation ($\sim$30~deg);
and the target is barely visible in individual frames.
We stacked 20 or 40 successive images with exposure times of 5~s,
and obtained images with effective exposure times of 100~s or 200~s.
A typical readout time of the CMOS sensors on TriCCS is 0.4~milliseconds,
which is negligibly small compared to the exposure time of 5~s.
We also stacked images using the World Coordinate System (WCS) of images 
corrected with the surrounding sources to suppress the elongations of the images of reference stars
(hereafter referred to as the sidereally stacked image).

We derived colors and magnitudes of \PT following the same procedure used in \cite{Beniyama2023a, Beniyama2023b, Beniyama2023c, Beniyama2024}.
Cosmic rays were removed with the \texttt{Python} package \texttt{astroscrappy} \citep{McCully2018}
using the Pieter van Dokkum's \texttt{L.A.Cosmic} algorithm \citep{vanDokkum2001}.
The circular aperture photometry was performed for \PT  using the SExtractor-based \texttt{Python} package \texttt{sep}. 
The aperture radii were set to \rad~pix, which is 
about 1.5–2.0 times as large as the FWHMs of the PSFs of the reference stars on the sidereally stacked images,
and annuli from \annin to \annout~pix from the center were used to estimate the background level and noise for \PT.
The circular aperture photometry was performed for the reference stars using \texttt{sep}, 
with a circular aperture after global background subtraction.
The aperture radii were set to \rad~pix as the photometry of \PT.
The photometric results of \PT and reference stars were obtained from the nonsidereal and sidereal stacked images, respectively.

All images were calibrated using Pan-STARRS catalog Data Release 2 \citep[DR2,][]{Chambers2016}.
Reference stars that met any of the following criteria were excluded from the analysis:
uncertainties in the $g$, $r$, $i$, or $z$ band magnitudes in the catalog larger than 0.05 mag; 
$(g-r)_{\mathrm{PS}}$ > 1.1; 
$(g-r)_{\mathrm{PS}}$ < 0.0; 
$(r-i)_{\mathrm{PS}}$ > 0.8; 
or $(r-i)_{\mathrm{PS}}$ < 0.0, 
where $(g-r)_{\mathrm{PS}}$ and $(r-i)_{\mathrm{PS}}$ are colors in the Pan-STARRS system.
Photometric measurements within 100 pixels from the edges of the image 
or contaminated by nearby sources within the aperture were excluded on a frame-by-frame basis.
Extended sources, possible quasars as well as variable stars were removed 
using \texttt{objinfoflag} and \texttt{objfilterflag} in the Pan-STARRS catalog.
Typically a few dozen of reference stars are used in each frame.
When the number of stars in each frame is less than five,
we do not use that frame to avoid any systematics in its color and/or magnitude estimates.
The typical uncertainties in magnitude zero points are less than 0.01~mag.

\subsection{Keck telescope}
Observations of 2024~PT$_5$ were obtained at the Keck Observatory with the MOSFIRE instrument 
on the Keck I 10-m telescope on Maunakea on 2025 January 16 under program R332 (PI: Bolin) and N085 (PI: Bolin).  
MOSFIRE has a 6.12\arcmin x 6.12\arcmin field of view, plate scale of 0.18~arcsec/pixel, 
and possesses a 
Y filter (central wavelength 1.048~$\mu$m, FWHM 0.152~$\mu$m), 
J filter (central wavelength 1.253~$\mu$m, FWHM 0.200~$\mu$m), 
H filter (central wavelength 1.637~$\mu$m, FWHM 0.341~$\mu$m), 
and K filter (central wavelength 2.147~$\mu$m, FWHM 0.314~$\mu$m) \citep[][]{McLean2012}. 

On 2025 January 16 UTC, 2024~PT$_5$ had a phase angle of 18.5
deg
, a geocentric distance of 0.013~au, and a heliocentric distance of 0.996~au, 
and on 2025 January 17, 2024~PT$_5$, had a phase angle of 20.6
deg
, a geocentric distance of 0.013~au, and a heliocentric distance of 0.996~au. 
The apparent sky motion of \PT was about 0.12~arcsec~s$^{-1}$.
The seeing measured by using in-field stars was $\sim$0.9~arcsec in the Y, J, and H bands on 2025 January 16 and $\sim$0.6~arcsec in the Y and K bands on 2025 January 17. 
The asteroid was observed in the Y, J, and H filters on 2025 January 16, 
and in the Y, and K filters on 2025 January 17 (Fig. \ref{fig:cutout_NIR}). 
Exposure times of 30~s, a five-point dither pattern was used, and the telescope was tracked at the rate of motion of the asteroid. 
Filters were changed between groups of five exposures between the Y, J, and H filters on 2025 January 16, and between the Y and K filters on 2025 January 17 to mitigate lightcurve effects on the photometry.
Calibration was completed by using in-field solar-like stars from the Pan-STARRS and 2MASS catalogs \citep[][]{Skrutskie2006,Tonry2012}.
Around 10 stars were used per image depending, and catalogue uncertainties for the standard stars in Y, J, H, and K bands ranged between 0.02-0.09 mag.

\begin{table*}
\caption{\label{t7}Summary of the observations\label{tab:obs}}
\centering
\begin{tabular}{rcccccccccc}
\hline\hline
Obs. Date & Tel. & Filter & $t_\mathrm{exp}$ & $N_\mathrm{img}$ & V     & $\alpha$ & $v$                & Air Mass   & Seeing   \\
(UTC)     &      &        & (s)              &                  & (mag) & (deg)    & (arcsec s$^{-1}$)  &            & (arcsec) \\
\hline
2025 Jan 04 15:23:43--16:02:39& Seimei & $g,r,i$ & 100 & 22 & 19.1 & 26.8 & 0.13 & 1.04--1.04 & 2.7\\
16:14:35--16:53:00& Seimei & $g,r,z$ & 100 & 19 & 19.1 & 26.7 & 0.14 & 1.05--1.07 & 2.8\\
17:01:18--17:39:44& Seimei & $g,r,i$ & 100 & 20 & 19.1 & 26.7 & 0.14 & 1.08--1.13 & 2.9\\
2025 Jan 07 16:49:28--17:47:58& Seimei & $g,r,i$ & 100 & 29 & 18.8 & 19.4 & 0.15 & 1.08--1.20 & 3.3\\
2025 Jan 10 15:13:50--16:00:38& Seimei & $g,r,z$ & 200 & 14 & 18.7 & 14.4 & 0.14 & 1.01--1.05 & 3.0\\
16:07:56--16:54:44& Seimei & $g,r,z$ & 200 & 10 & 18.7 & 14.4 & 0.14 & 1.06--1.14 & 3.1\\
2025 Jan 16 08:44:49--08:47:29& Keck I & Y & 30 & 5 & 18.9 & 18.5 & 0.12 & 1.01--1.01 & 0.9\\
08:50:56--08:53:53& Keck I & J & 30 & 5 & 18.9 & 18.5 & 0.12 & 1.00--1.00 & 0.9\\
08:55:33--08:58:17& Keck I & Y & 30 & 5 & 18.9 & 18.5 & 0.12 & 1.00--1.00 & 0.9\\
08:59:38--09:05:20& Keck I & H & 30 & 5 & 18.9 & 18.5 & 0.12 & 1.00--1.00 & 0.9\\
2025 Jan 17 09:13:43--09:16:26& Keck I & Y & 30 & 5 & 19.0 & 20.6 & 0.12 & 1.00--1.00 & 0.6\\
09:24:17--09:27:58& Keck I & K & 30 & 5 & 19.0 & 20.6 & 0.12 & 1.00--1.01 & 0.6\\
\hline
\end{tabular}
\tablefoot{
Observation time in UT in midtime of exposure (Obs. Date), telescope (Tel.), filters (Filter), 
total exposure time per frame ($t_{\mathrm{exp}}$),
and the number of images ($N_\mathrm{img}$) are listed.
Predicted V band apparent magnitude (V), 
phase angle ($\alpha$),
and 
apparent angular rate of 2024 PT$_{5}$ ($v$)
at the observation starting time
are referred to NASA Jet Propulsion Laboratory (JPL) Horizons
as of July 08, 2025.
Elevations of 2024 PT$_{5}$ to calculate air mass range (Air Mass) are 
also referred to NASA JPL Horizons.
Seeing FWHM (Seeing) in the r band for Seimei observations and in the Y, J, H, and K bands for Keck observations
measured by computing the FWHM of reference stars are also listed.
}
\end{table*}

\section{Results} \label{sec:result}
\subsection{
Lightcurves, rotation period, and axial ratio
} \label{subsec:res_lc}
\begin{figure*}
\centering
\includegraphics[width=1.0\hsize]{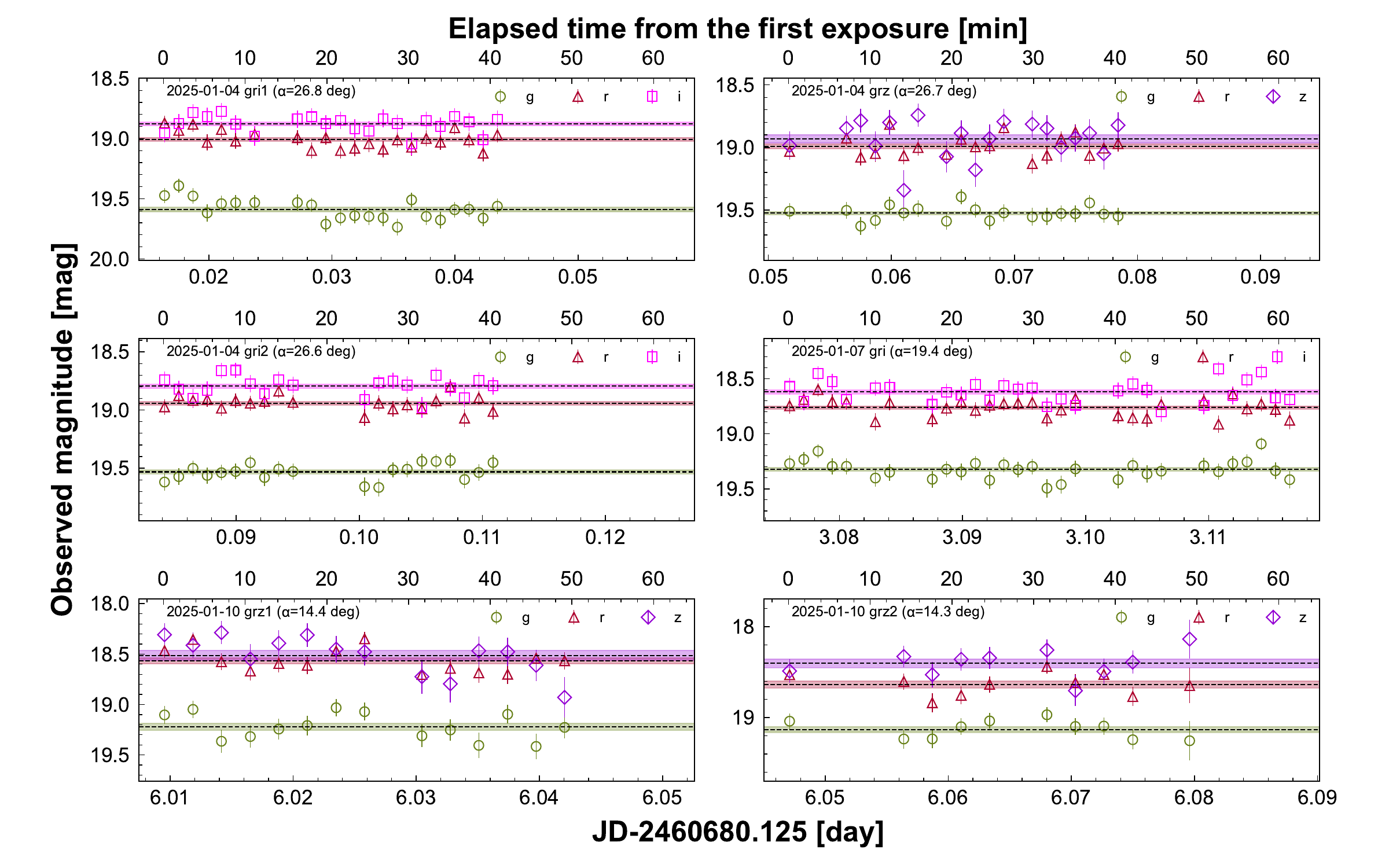}
\caption{
Lightcurves of \PT.
The observed $g$, $r$, $i$, and $z$ bands magnitudes 
are presented as circles, triangles, squares, and diamonds, respectively.
Bars indicate the 1$\sigma$ uncertainties.
The arithmetic average of magnitude in each light curve is presented with a dashed line. 
Shaded areas indicate the standard errors of the averaged magnitudes.
}
\label{fig:lc}
\end{figure*}

\begin{figure*}
\centering
\includegraphics[width=1.0\hsize]{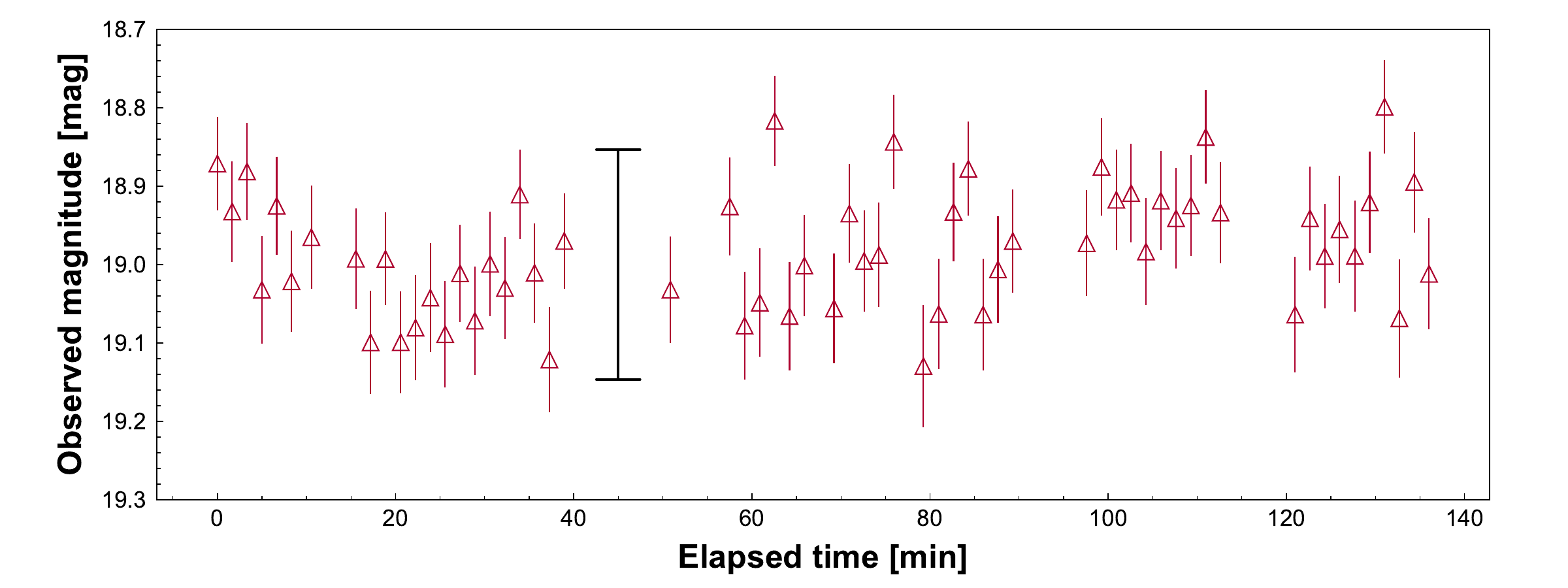}
\caption{
Lightcurves of \PT in the r-band on January 4.
Bars indicate the 1$\sigma$ uncertainties.
The variation of lightcurve, 0.3~mag, is indicated.  
}
\label{fig:lc_r}
\end{figure*}

The lightcurves of \PT are shown in Fig. \ref{fig:lc}.
We see variations with amplitudes of approximately 0.3~mag in all lightcurves.
The lightcurves obtained on 2025 January 4 has an observation arc longer than 2~hrs,
and the corresponding lightcurve in the $r$ band are presented in Figure \ref{fig:lc_r}.
This is longer than any existing observations (see Table \ref{tab:existingobs}).

\begin{figure}
\centering
\includegraphics[width=1.0\hsize]{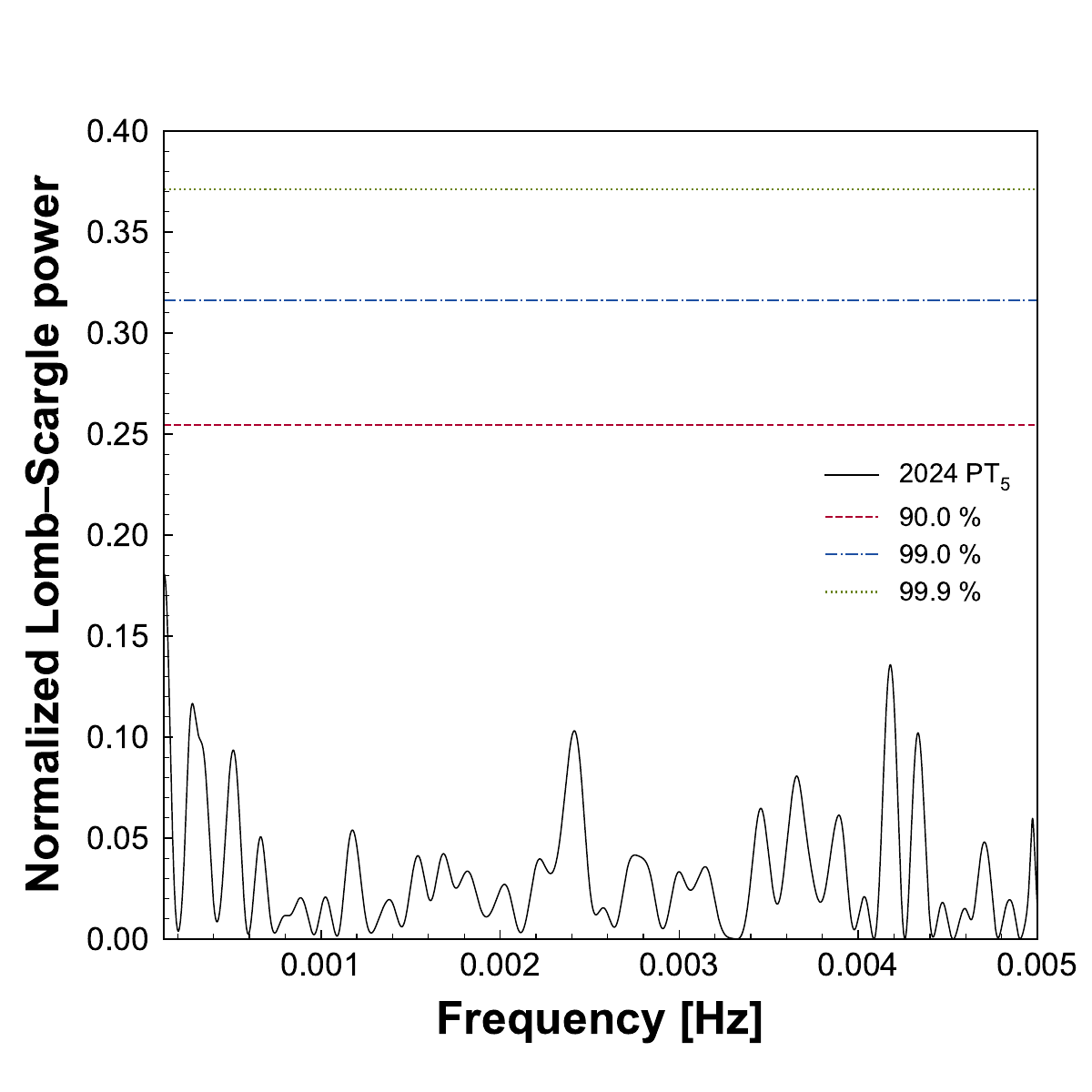}
\caption{
Lomb--Scargle periodogram of \PT.
The number of harmonics of the model curve is unity.
Dashed, dot-dashed, and dotted horizontal lines show 90.0, 99.0, and 99.9\% confidence levels, respectively.
}
\label{fig:LS}
\end{figure}

We performed the periodic analysis with the three $r$-band lightcurves obtained on 2025 January 4 
using the Lomb--Scargle technique \citep{Lomb1976, Scargle1982, VanderPlas2018}.
The Lomb--Scargle periodograms with a period range between \Prmin~s to \Prmax~s are shown in Fig. \ref{fig:LS}.
We showed 90.0, 99.0, and 99.9\% confidence levels in the periodogram.
The minimum and maximum of the period range correspond to twice 
the sampling rate and observation arc, respectively.
We found no periodicity in our lightcurves.

We assumed the asteroid is a triaxial ellipsoid with
axial lengths of $A$, $B$, and $C$ ($A \geq B \geq C$) and the aspect angle, angle between rotation axis and the asteroids--observer direction, of 90~deg. 
A lower limit of axial ratio $A/B$ is estimated as follows:
\begin{equation}
A/B \geq 10^{0.4 m(\alpha)/(1+s\alpha)},
\end{equation}
where $m(\alpha)$ is the lightcurve amplitude 
at a phase angle of $\alpha$ and $s$ is a slope depending on the taxonomic type of the asteroid \citep{Bowell1989}.
When we assume $s$ of 0.030, a typical value of S-type asteroids \citep{Zappala1990}, and a lightcurve amplitude of 0.3, 
axial ratio $A/B$ of \PT is larger than 1.17.

\subsection{Colors} \label{subsec:res_col}
The 
averages and standard deviations of the mean colors of \PT calculated for each observing block
were derived as \gr, \ri, and \rz. 
We summarize the derived colors of \PT and 
with those reported in previous studies in Table \ref{tab:col}.
We converted colors from \cite{Bolin2025_PT5} in the SDSS system into the Pan-STARRS system using the following equations \citep{Tonry2012}:
\begin{eqnarray}
g &=& g_\mathrm{SDSS} - 0.012 - 0.139(g_\mathrm{SDSS}-r_\mathrm{SDSS}), \\
r &=& r_\mathrm{SDSS} + 0.000 - 0.007(g_\mathrm{SDSS}-r_\mathrm{SDSS}), \\
i &=& i_\mathrm{SDSS} + 0.004 - 0.014(g_\mathrm{SDSS}-r_\mathrm{SDSS}), \\
z &=& z_\mathrm{SDSS} - 0.013 + 0.039(g_\mathrm{SDSS}-r_\mathrm{SDSS}), 
\end{eqnarray}
where $g$, $r$, $i$, and $z$ are magnitudes in the Pan-STARRS system,
while 
$g_\mathrm{SDSS}$, $r_\mathrm{SDSS}$, $i_\mathrm{SDSS}$, and $z_\mathrm{SDSS}$ 
are magnitudes in the SDSS system.
We computed the propagated uncertainties of the colors in the Pan-STARRS system
with the photometric errors and uncertainties in conversions.
Table \ref{tab:col} also includes colors of 
\Kamo from \citet[][]{Sharkey2021}
and 
lunar rock core samples from \citet[][]{Isaacson2011}.
We used the techniques from \citet[][]{Bolin2021, Bolin2022} to estimate the visible colors in the Pan-STARRS system from spectra.
We used the spectroscopy module from \texttt{sbpy}, 
an astropy-affiliated package for small-body planetary astronomy \citep{Mommert2019}, 
and solar colors from \citep[][]{Willmer2018}.

\begin{table*}
\caption{Colors in the Pan-STARRS system\label{tab:col}}
\centering
\begin{tabular}{lllll}
\hline\hline
Objects                       & Reference             & $g-r$ & $r-i$ & $r-z$    \\
\hline
\PT                           & \cite{Bolin2025_PT5}  & \grBolinPS & \riBolinPS & \rzBolinPS \\ 
& \cite{Kareta2025}     & \grKaretaPS & \riKaretaPS & \rzKaretaPS \\         
& This study            & \grval   & \rival & \rzval \\
\hline
\Kamo                         & \cite{Sharkey2021}    & \grKamoPS & \riKamoPS & \rzKamoPS \\
Lunar rock core samples       & \cite{Isaacson2011}   & \grlunarPS & \rilunarPS & \rzlunarPS \\
\hline
\end{tabular}
\tablefoot{
Colors of \PT in the SDSS system reported in \citet{Bolin2025_PT5} were converted to those in the Pan-STARRS system using the equations in \cite{Tonry2012}.
}
\end{table*}

\begin{figure}
\centering
\includegraphics[width=1.0\hsize]{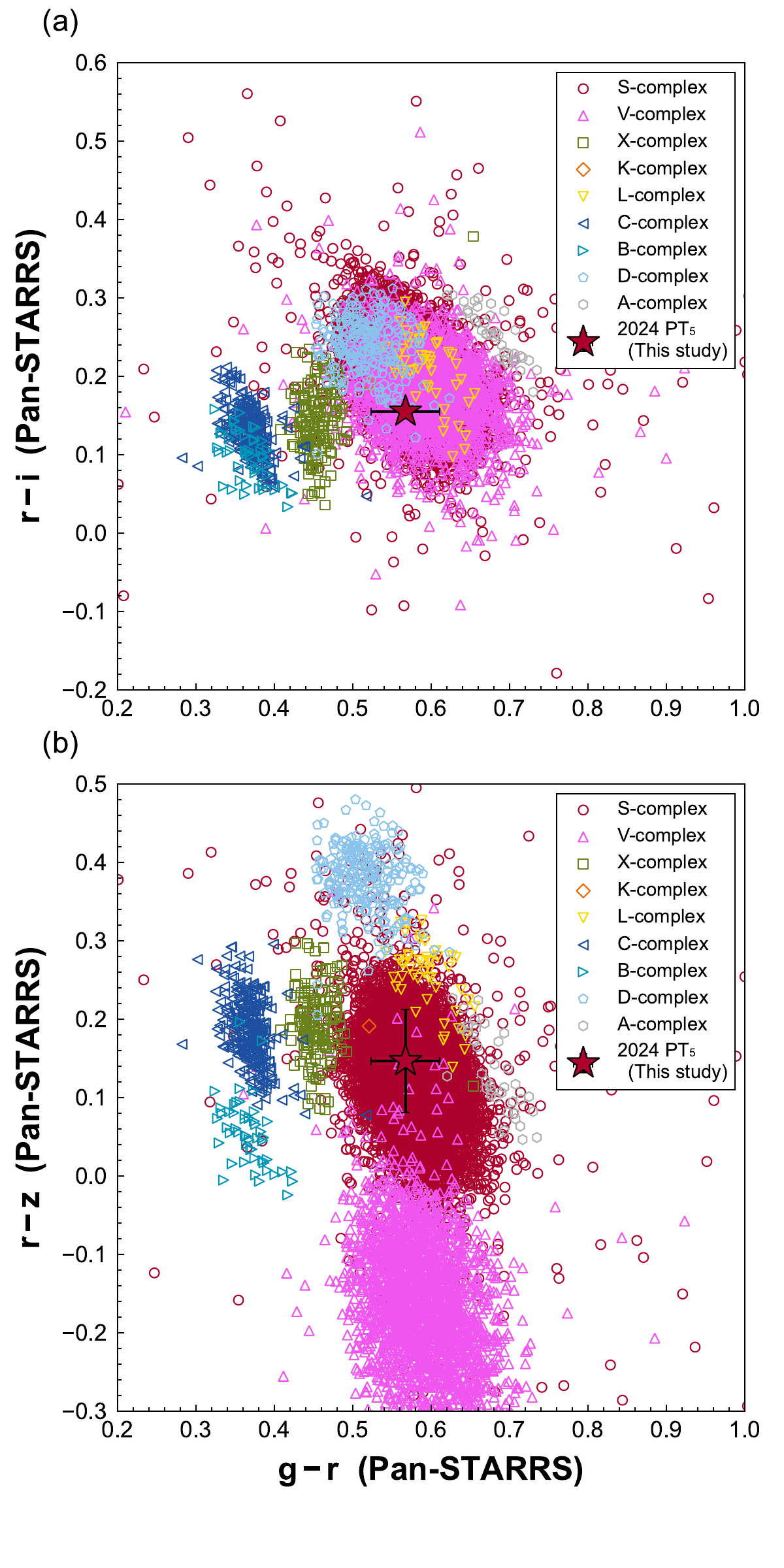}
\caption{
Color-color diagram of (a) $g$-$r$ vs. $r$-$i$ and (b) $g$-$r$ vs. $r$-$z$.
Mean of the individual nightly mean colors of \PT are plotted with stars. 
Error bars indicate the standard deviation of nightly mean colors.
Error bars indicate the 1$\sigma$ uncertainties.
Asteroids from \citep{Sergeyev2021} are also plotted: 
S-complex (circles), 
V-complex (triangles),
X-complex (squares),
K-complex (diamonds),
L-complex (inverse triangles),
C-complex (left-pointing triangles),
B-complex (right-pointing triangles),
D-complex (pentagons),
and 
A-complex (hexagons). 
}
\label{fig:cc}
\end{figure}

We present the derived colors of \PT in Fig. \ref{fig:cc} 
with those of asteroids from the recent catalog \citep{Sergeyev2021}.
In this catalog, the probabilities of the asteroid complex are assigned for each asteroid. 
We extracted asteroids with 80\% or higher probabilities to belong to a specific complex except \texttt{U}, which indicates an unknown class.
We calculated $g-r$, $r-i$, and $r-z$ in the SDSS system, 
and then converted these colors into the Pan-STARRS system using equations given in \citet{Tonry2012}.
By a visual inspection, 
\PT is overlapping with S-complex asteroids and lunar samples in the color-color diagrams.

\begin{figure}
\centering
\includegraphics[width=1.0\hsize]{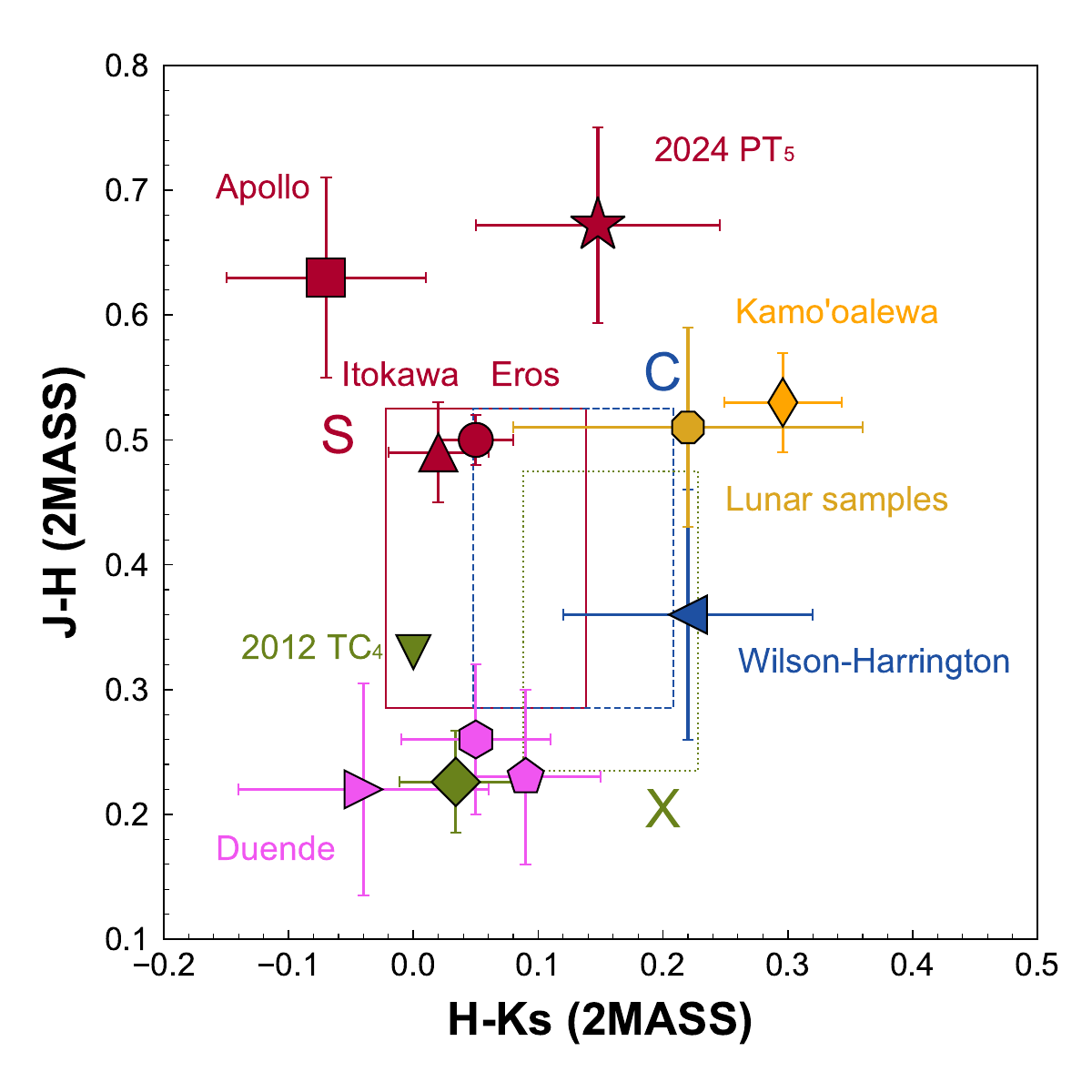}
\caption{
Color-color diagram of $J$-$H$ vs. $J$-$Ks$.
Colors of \PT are plotted with a star.
Colors of near-Earth asteroids in the literature are plotted:
Eros (circle) \citep{Chapman1976},
Itokawa (triangle) \citep{Ishiguro2003},
Apollo (square) \citep{Hartmann1982},
2012~TC$_4$ (diamond and inverse triangle) \citep{Urakawa2019, Reddy2019},
Wilson-Harrington (left-pointing triangle) \citep{Hartmann1982},
and Duende (right-pointing triangle and pentagon) \citep{deLeon2013, Takahashi2014}.
Error bars indicate the 1$\sigma$ uncertainties.
Average near-infrared colors of S-, C-, and X-types observed by 
VISTA-VHS are also indicated with rectangles \citep{Popescu2016}.
Colors of lunar rock core samples \citep{Isaacson2011} and 
\Kamo \citep{Sharkey2021} are also shown 
by octagon and elongated diamond, respectively. 
The error bars correspond to the uncertainties of 1$\sigma$ envelope of the compiled spectra in Fig. \ref{fig:ref}.
}
\label{fig:nircc}
\end{figure}

We presented the derived near-infrared colors of \PT in Fig. \ref{fig:nircc}. 
We also plotted 
near-infrared
colors of NEAs from the literature:
S-type (433)~Eros \citep{Chapman1976},
S-type (25143)~Itokawa \citep{Ishiguro2003},
S-type (1862)~Apollo and C-type (4015)~Wilson-Harrington \citep{Hartmann1982},
X-type 2012TC$_4$ \citep{Urakawa2019, Reddy2019}.
and L-type (367943)~Duende \citep{deLeon2013, Takahashi2014}.
Solar system small bodies are serendipitously observed by
survey observations in the near-infrared using 4.1~m 
Visible and Infrared Survey Telescope for Astronomy (VISTA) at ESO's Cerro Paranal Observatory, Chile.
Infrared colors of minor bodies are extracted in \cite{Popescu2016, Popescu2018}.
We presented the infrared colors of S, C, and X-type asteroids in Fig. \ref{fig:nircc} for comparison.
We computed the $Y$-$J$, $Y$-$H$, and $J$-$H$ colors using the $Y$, $J$, and $H$ observations from 2025 Jan 16. We compute the $Y$-$K_s$ color of \PT using the $Y$ and $K_s$ observations from 2025 Jan 17.
The near-infrared colors of \PT are comparable to 
those of a Q-type asteroid Apollo and lunar samples, and $J$-$H$ color is redder than those of S-types, Eros, and Itokawa. The $Y$-$J$ color of \PT is estimated to be \YJbyBB. To compute the $H$-$K_s$ color of \PT, we subtract the $Y$-$H$ color of \PT based on the $Y$ and $H$ observations taken on 2025 Jan 16 from the $Y$-$K_s$ color obtained from the $Y$ and $K_s$ observations 2025 Jan 17. 
The 
near-infrared
color of \PT is similar to that of small S-complex NEA 2024~YR$_4$, measured with the same instruments \citep{Bolin2025_YR4}, and consistent with its $J$-$H$ and $H$-$K_s$ colors. However, we identified possible systematics in the $Y$–$J$ color, likely caused by differences between the MOSFIRE and Pan-STARRS $Y$-band filter response functions. Therefore, we exclude the $Y$–$J$ color of \PT in the subsequent analysis.

\subsection{Phase curve} \label{subsec:res_pc}
\begin{figure}
\centering
\includegraphics[width=1.0\hsize]{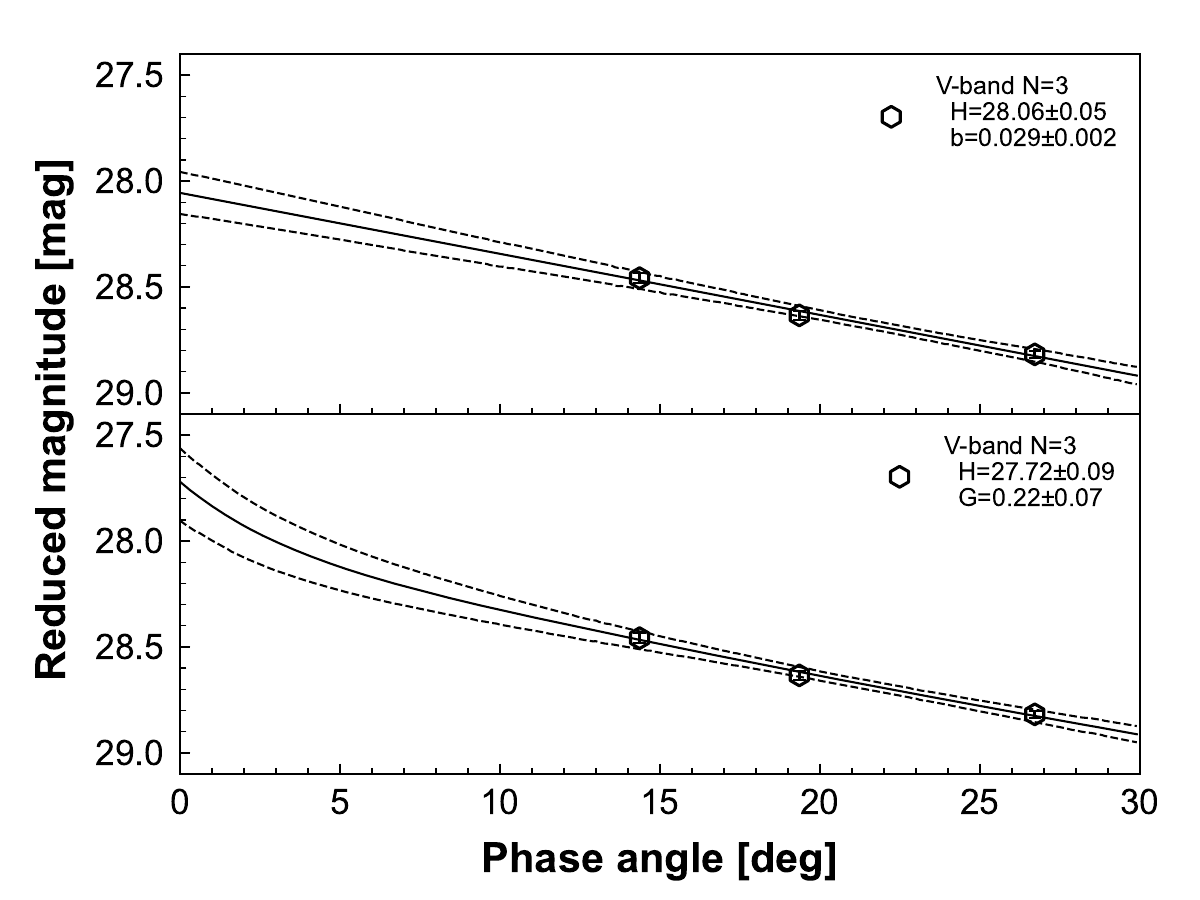}
\caption{
Phase angle dependence of reduced $V$ magnitudes of \PT.
Bars indicate the 1$\sigma$ uncertainties.
All magnitudes obtained on the same day are averaged and plotted. 
(top)
The medians (50th percentile) of fitting model curves with the linear model are presented as solid lines. 
(bottom)
The medians (50th percentile) of fitting model curves with the $H$-$G$ model are presented as solid lines. 
Uncertainty envelopes representing the 95\% highest density interval values are shown by dashed lines.
}
\label{fig:pc}
\end{figure}

We observed \PT across a wide phase angle range, from \alphamin to \alphamax.
The Pan-STARRS magnitudes in the $g$ and $r$ bands were converted to Johnson $V$-band magnitudes using the 
following transformation provided by \citet{Tonry2012}:
\begin{equation}
V = r + 0.006 + 0.474(g-r).
\end{equation}
The phase curves of \PT in the $V$-band are shown in Fig. \ref{fig:pc}.

Since our phase curve lacks observations at lower phase angles, 
where opposition surges appear, we fitted it with simple models.
We derived an absolute magnitude and slope of the phase curve with the linear model:
\begin{equation}
H_V(\alpha) = H_{V, \mathrm{linear}} + b\alpha,
\end{equation}
where $H_{V, \mathrm{linear}}$ is an absolute magnitude in the $V$-band 
and $b$ is the slope of the fitting curve.
We also derived an absolute magnitude and slope of the phase curve with the $H$-$G$ model \citep{Bowell1989}:
\begin{equation}
H_V(\alpha) = H_{V, HG} - 2.5 \log_{10}{((1-G)\Phi_1(\alpha)+G\Phi_2(\alpha)}),
\end{equation}
where $H_{V, HG}$ is an absolute magnitude in the $V$-band 
and $G$ is the slope of the fitting curve.
$\Phi_1$ and $\Phi_2$ are phase functions written as follows with a basic function $W$:
\begin{align}
\Phi_1(\alpha) &= W\left(1-\frac{0.986\sin{\alpha}}{0.119+1.341\sin{\alpha}-0.754\sin^2{\alpha}}\right) \nonumber \\
&+ (1-W) \exp{\left(-3.332\tan^{0.631}{\frac{\alpha}{2}}\right)},\\
\Phi_2(\alpha) &= W\left(1-\frac{0.238\sin{\alpha}}{0.119+1.341\sin{\alpha}-0.754\sin^2{\alpha}}\right) \nonumber \\
&+ (1-W) \exp{\left(-1.862\tan^{1.218}{\frac{\alpha}{2}}\right)},\\
W              &= \exp{\left(-90.56\tan^2{\frac{\alpha}{2}}\right)}.
\end{align}
The uncertainties of $H_{V, \mathrm{linear}}$, $b$, $H_{V, HG}$, and $G$ were estimated with the Monte Carlo technique. 
We made 3000 phase curves by randomly resampling the data by assuming each observed datum follows a normal distribution whose standard deviation is the standard error of the 
error-weighted average
magnitude.
We derived fitting parameters as follows:
$H_{V, \mathrm{linear}}$ = \HVlinear,
$b$ = \bVlinear,
$H_{V, HG}$ = \HVHG,
and
$G$ = \GV.

\begin{figure}
\centering
\includegraphics[width=1.0\hsize]{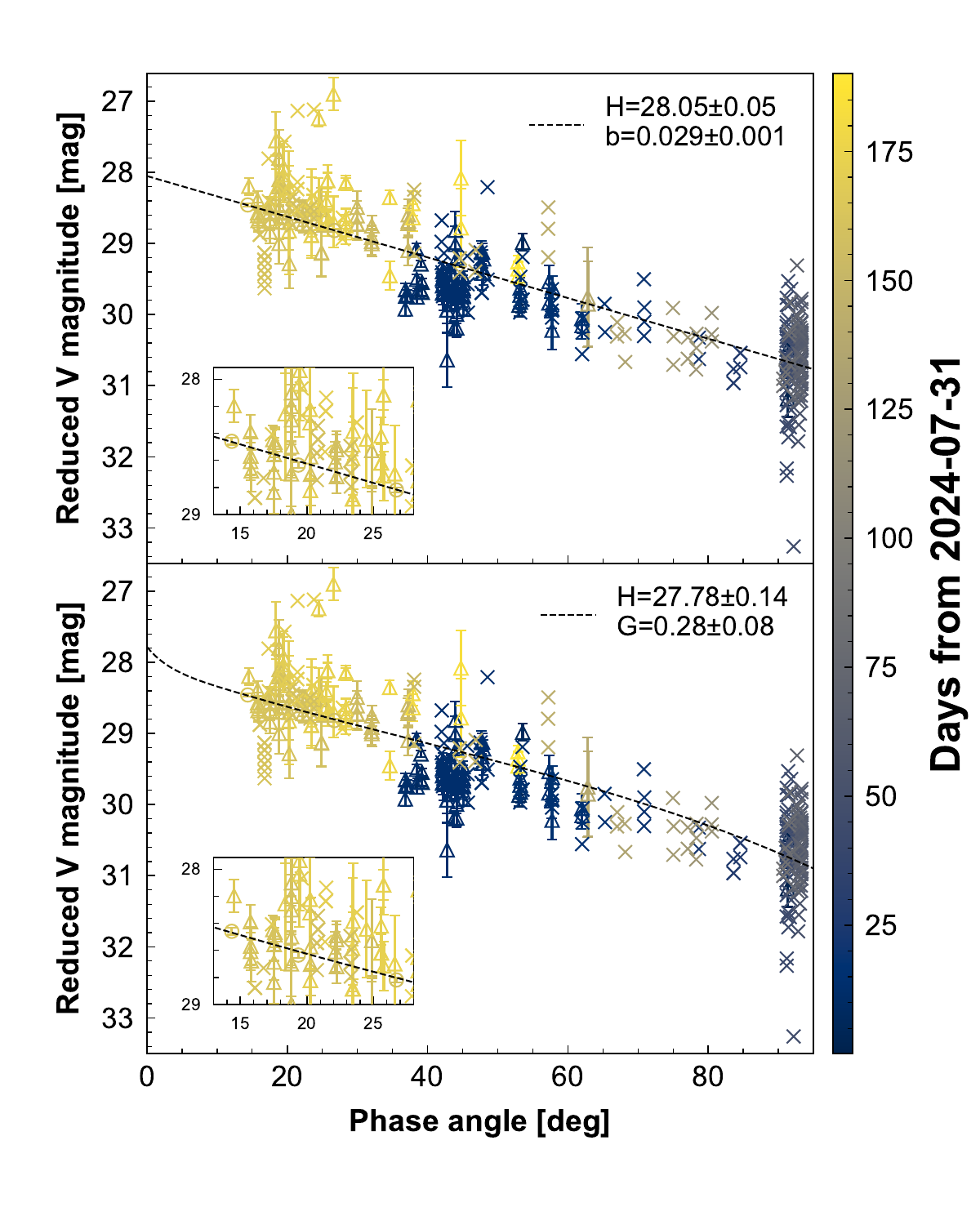}
\caption{
Phase angle dependence of reduced $V$-band magnitudes of \PT, including corrected observations from the MPC database.
(top)
Magnitudes obtained on the same day from Seimei/TriCCS are averaged and plotted by circles.
Data from the MPC database with reported magnitude uncertainties are shown as triangles, 
while those without reported uncertainties are represented by crosses. 
Bars indicate the 1$\sigma$ uncertainties.
The fitting model curve with the linear model is presented as a dashed line.
Each data point is color-coded by the observation date, expressed as days since 2024 July 31.  
The figure includes an inset plot that magnifies a region including data from Seimei/TriCCS. 
(bottom)
Same as the top panel, but the data are fitted with the $H$-$G$ model.
}
\label{fig:pc_wmpc}
\end{figure}

The Minor Planet Center (MPC) database contains 410 observations of \PT obtained from 44 different observatories. 
Of these, 404 observations include reported magnitudes. We used this dataset to estimate the phase curve of \PT.
To merge the heterogeneous data into a single reference photometric system, we collected observations of various 
asteroids from the MPC that were obtained using the same observatories and filters. 
We then computed the differences between the reported and predicted V magnitudes for these reference asteroids. 
The predicted V magnitudes were obtained using the \texttt{Miriade} software. 
These differences were used to correct the reported magnitudes of \PT to build a consistent dataset
as shown in Fig. \ref{fig:pc_wmpc}.

The fitting yielded the following parameters:
$H_{V, \mathrm{linear}}$ = \HVlinearMPC,
$b$ = \bVlinearMPC,
$H_{V, HG}$ = \HVHGMPC,
and
$G$ = \GVMPC.
The uncertainties in the parameters are given by the 1$\sigma$ 
calculated as the square roots of the diagonal elements of the covariance matrix.
These values are consistent with those derived from our own measurements, however, the associated uncertainties are larger. 
Therefore, we adopt the parameters obtained solely from our measurements in the following discussion.
We note that our computation is based on only three points, 
while the MPC data includes 404 points. 
Nevertheless, the results are very similar, 
and the consistency in results confirms the reliability of our values.

\section{Discussion} \label{sec:disc}
\subsection{Physical properties}
The rotation state of \PT is still not clear, even with our lightcurves throughout more than 2~hr. 
The possibility that these variations are due to noise cannot be excluded.
However, all published lightcurves, including ours, exhibit variations with a period of several tens of minutes.
This indicates a tumbling motion of \PT, as noted in previous studies \citep{Kareta2025, Marcos2025}. 
The fact that \PT is a tumbling asteroid supports the lunar ejecta origin 
since tumbling motion is expected after the collisional event \citep{Harris1994}.
We note that the tidal force can change the rotation state of \PT during the close approach, as is reported for another tiny 
NEA (367943)~Duende \citep[a.k.a. 2012~DA$_{14}$,][]{Moskovitz2020, Benson2020}.
\PT experienced a close approach with a distance of 0.00379~au or 1.5 Lunar distances (LD) from the Earth on 2024 August 8.
The close encounter on 2024 August 8 and/or earlier close encounters may have changed the spin state of \PT inducing a tumbling state.
This effect will be well studied for (99942)~Apophis in 2029 during its $\sim0.07$~LD close approach \citep{Ballouz2024}.

\begin{figure}
\centering
\includegraphics[width=1.0\hsize]{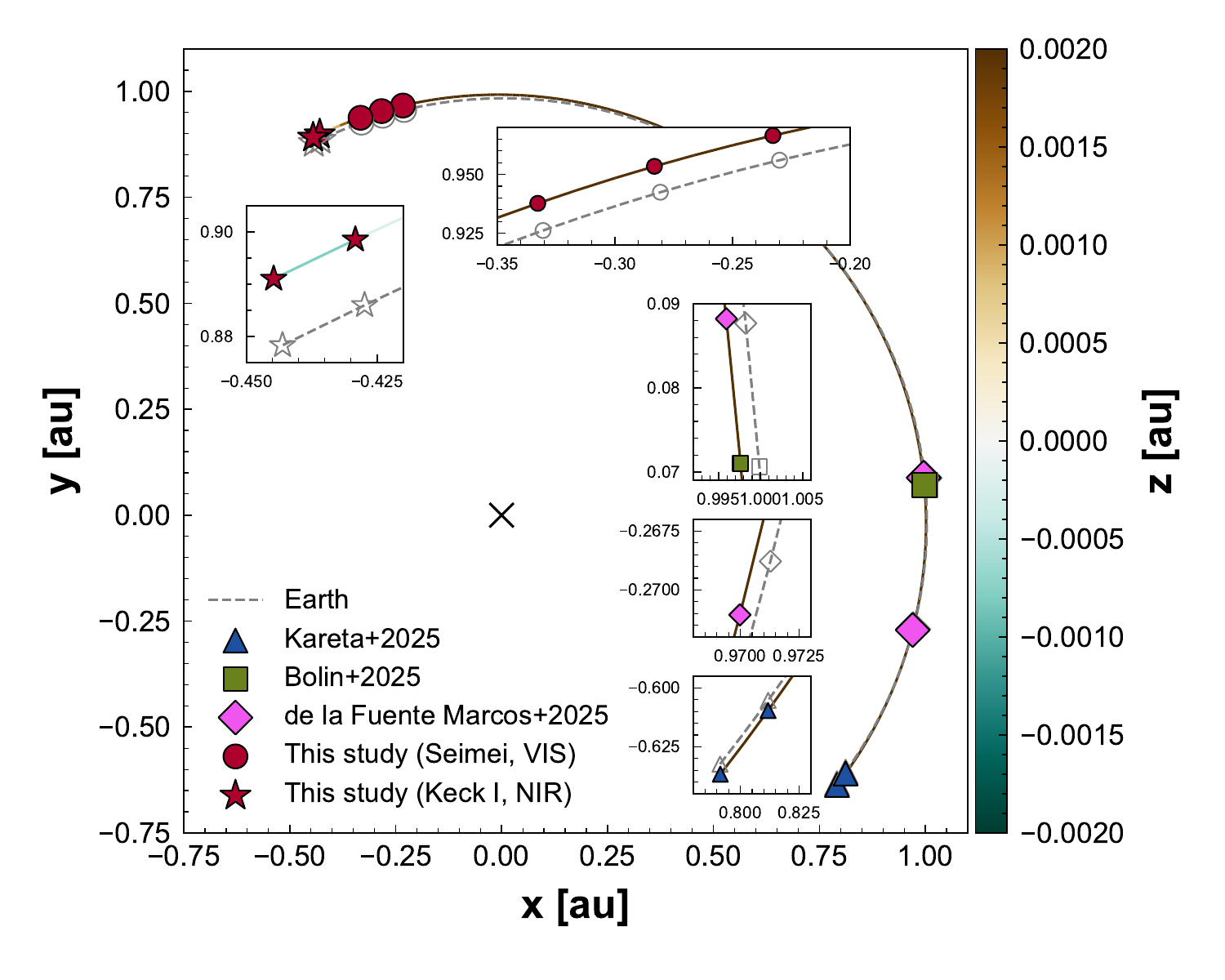}
\caption{
Heliocentric positions of \PT and Earth at the time of observations
in \cite{Kareta2025}, \cite{Bolin2025_PT5}, \cite{Marcos2025}, and this study.
Positions of \PT and Earth are indicated by filled markers and open markers, respectively.
Orbits of \PT and Earth are shown by solid and dashed lines, respectively. 
The color of the solid line indicates the 
$z$-coordinate (vertical position) of \PT in the heliocentric frame.
The figure includes multiple inset plots that magnify specific regions.
Although their scales differ, the aspect ratio is kept constant.
}
\label{fig:loc}
\end{figure}

The derived visible colors of \PT 
in Table \ref{tab:col} and
Fig. \ref{fig:cc} are largely consistent with 
previous measurements by \cite{Bolin2025_PT5} and \cite{Kareta2025}, but with some differences 
of about 0.1 mag at most.
This can be 
qualitatively
explained when \PT's surface is not homogeneous within the uncertainties of measurements.
Since our observation geometry was dramatically different from other studies as shown in Fig. \ref{fig:loc}, 
we might observe a different side of \PT compared to observations in 2024.
However, qualitatively it is unrealistic to detect a change in colors of about 0.1 mag unless 
roughly half of the \PT's surface would have significantly different composition compared to the other half.
A possible 
explanation is that \PT is tumbling, 
and the derived colors in previous observations are affected by the complex brightness variations.
The simultaneous multicolor photometry we performed is a reliable method,
unaffected by a possible complex rotation of \PT and less affected by observational artifacts.

As in the case of the rotation state, 
we note that close encounters with planets can change the surface properties, 
exposing fresh unweathered surface grains \citep[e.g.,][]{Binzel2010}.
Very recently, however, \cite{McGraw2024} observed 2024~MK, a S-complex NEA like \PT, 
before and after its close approach of 0.76~LD from the Earth. They obtained multiple high-quality 
visible to near-infrared
spectra of 2024~MK using IRTF/SpeX on Mauna Kea.
They found no planetary-encounter-induced spectral changes in the close approach, 
in contrast to previous works \citep[e.g.,][]{Binzel2010},
and concluded that a single close encounter at 0.76~LD may not be enough to cause a detectable change in ground-based observations. 
As in the case of 2024~MK in \cite{McGraw2024}, the phase reddening effect is also negligible for \PT since our observations are performed at phase angles less than \alphamax.

The $b$ parameter derived in Sect. \ref{subsec:res_pc} has a tight correlation with geometric albedo \citep{Belskaya2000}:
\begin{equation}
b = C_1 - C_2 \log_{10} p_V, \label{eq:bpV}
\end{equation}
where $C_1$ and $C_2$ are parameters obtained with phase slopes of asteroids whose albedo is estimated from thermal infrared observations.
We used updated parameters $C_1 = 0.016\pm0.001$ and $C_2 = 0.022\pm0.001$ from \cite{Shevchenko2021}.
High-albedo asteroids have a shallower slope (smaller $b$) 
in the phase curves since the contribution of shadowing decreases as albedo increases \citep{Belskaya2000}.
For instance, 
\cite{Shevchenko2012} estimated $b=0.040$--$0.045$~mag~deg$^{-1}$ for low-albedo Jupiter Trojans,
\cite{Ishiguro2014} estimated $b=0.039\pm0.001$~mag~deg$^{-1}$ for low-albedo NEA (162173)~Ryugu,
and 
\cite{Reddy2015} estimated $b=0.0225\pm0.0006$~mag~deg$^{-1}$ for high-albedo V-type NEA (357439)~2004~BL$_{86}$.

We derived a geometric albedo of \PT as \pVlinear with $b_V$ derived from only our measurements.
This is a fairly good match to S-type ($p_V=0.24^{+0.04}_{-0.05}$) and Q-type ($p_V=0.25^{+0.07}_{-0.04}$) NEAs \citep{Marsset2022a}.
This is confirmed from our $G_V$ of \GV, which is consistent with a 
typical value for S-types \citep{Warner2009}.  
The diameter of \PT can be estimated with our updated $H_{V,HG}$ and $p_V$ using the following
equation \citep{Fowler1992, Pravec2007}:
\begin{equation}
D=\frac{1329}{\sqrt{p_\mathrm{V}}}\times10^{-H_{V,HG}/5}. \label{eq:D}
\end{equation}
The diameter of \PT is estimated to be \DiamVHG~m, which is consistent with the previous estimates, 
$5.4\pm1.2$~m \citep{Bolin2025_PT5} and $8 \leq D \leq 12$~m \citep{Kareta2025}.

A caveat is that it is unclear if we can apply the empirical relationship 
of Eq. \ref{eq:bpV}
to such a tiny asteroid.
The empirical relation in \cite{Belskaya2000, Shevchenko2021} is based on the WISE and AKARI measurements of relatively large asteroids, 
and it may not apply to tiny asteroids like \PT if their surface properties are different from those of large asteroids.
\cite{Hasselmann2024} reported an unusual photometric phase curve of the NASA DART mission \citep{Rivkin2021, Terik2023} 
and the ESA HERA mission \citep{Michel2022} target S-type NEA (65803)~Didymos ($D\sim800$~m).
They intensively analyzed the photometric phase curve of Didymos obtained at a wide range of phase angles, focusing not only on the linear slope but also on the opposition effect.
They concluded that the photometric phase curve parameters are 
similar to those of M-type or C-complex asteroids, depending on the model used.
To quantitatively estimate albedo, cross validations by thermal infrared observations 
and/or polarimetry are important \cite[e.g.,][]{Selmi2025}. 
As for Didymos, it is expected that the Hera mission might end the discussion related to its surface properties.
The Hayabusa2 extended mission \citep{Hirabayashi2021} and Tianwen-2 mission \citep{Zhang2021} will also enhance
our understanding of two fast-rotating tiny NEAs, 1998~KY$_{26}$ ($D\leq40$~m) and \Kamo ($D\leq$100~m), respectively.

\subsection{Origin of 2024~PT$_5$}
Previous studies showed that the reflectance spectra of \PT match 
those of the lunar samples \citep{Bolin2025_PT5, Kareta2025, Marcos2025}.
\cite{Bolin2025_PT5} concluded that the physical properties of \PT
are compatible with an inner Main Belt or lunar ejecta with a preference to the latter,
while both \cite{Kareta2025} and \cite{Marcos2025} concluded that 
their reflectance spectra are suggestive of a lunar origin.
Their arguments are based on the similarity of reflectance spectra.
\cite{Marcos2025} compared their visible to near-infrared (0.475~$\mu$m to 0.880~$\mu$m) 
reflectance spectrum of \PT with template spectra from \cite{DeMeo2009}, and showed that the best-matching
template is Sv-type, while they compared the spectrum of \PT with over 15000 spectra of meteorites, terrestrial rocks, and lunar
soils in the RELAB database \citep{Pieters1983} using the M4AST tool \citep{Popescu2012}.
Similarly, \cite{Kareta2025} compared their visible to near-infrared ($\sim$0.40~$\mu$m to 2.45~$\mu$m)
reflectance spectrum of \PT with template spectra from \cite{DeMeo2009}, and concluded that no taxonomic class fits the spectrum of \PT.
They compared the spectrum of \PT with every sample in the RELAB database. 

\begin{figure*}[ht]
\centering
\begin{subfigure}[b]{0.45\hsize}
\includegraphics[width=\linewidth]{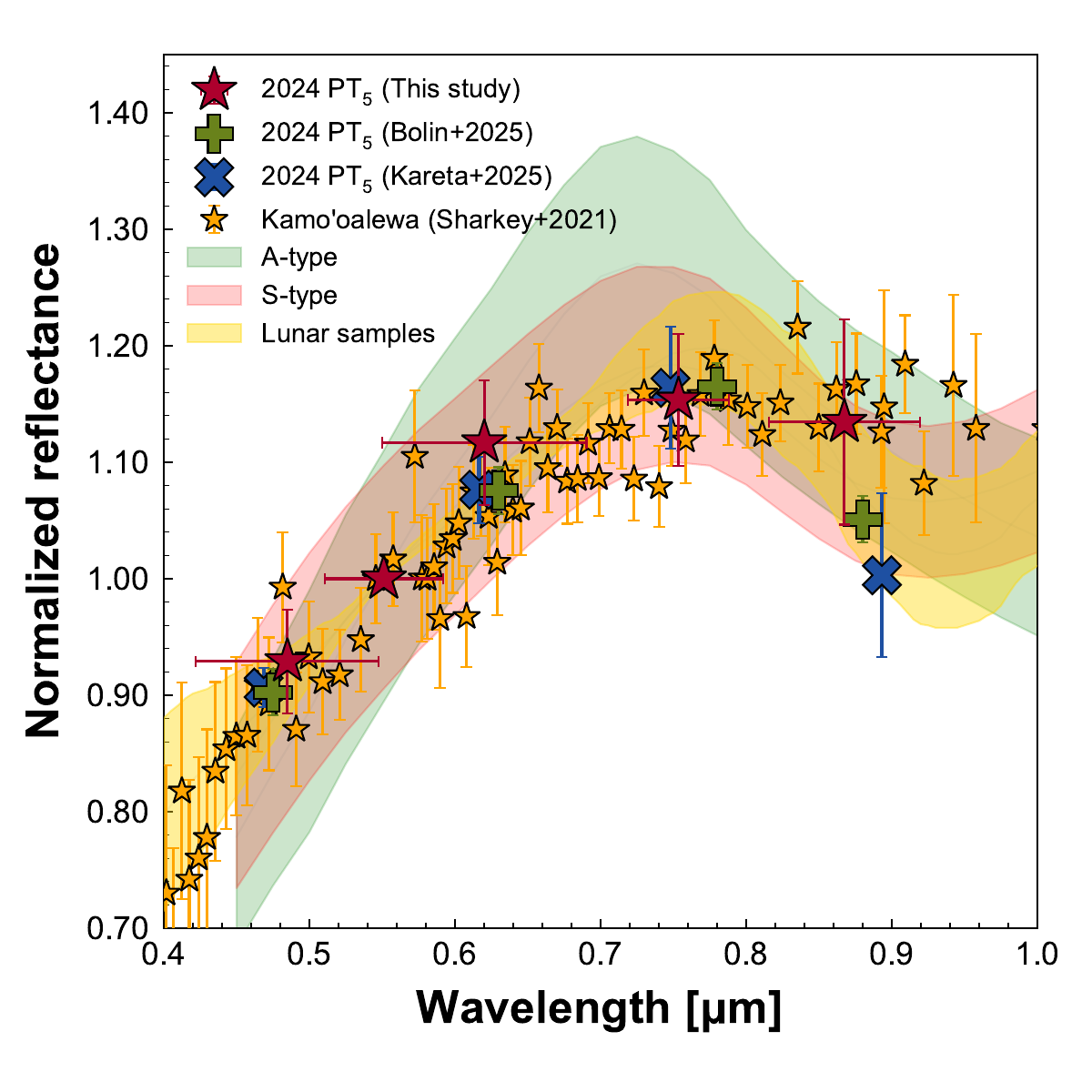}
\end{subfigure}
\begin{subfigure}[b]{0.45\hsize}
\includegraphics[width=\linewidth]{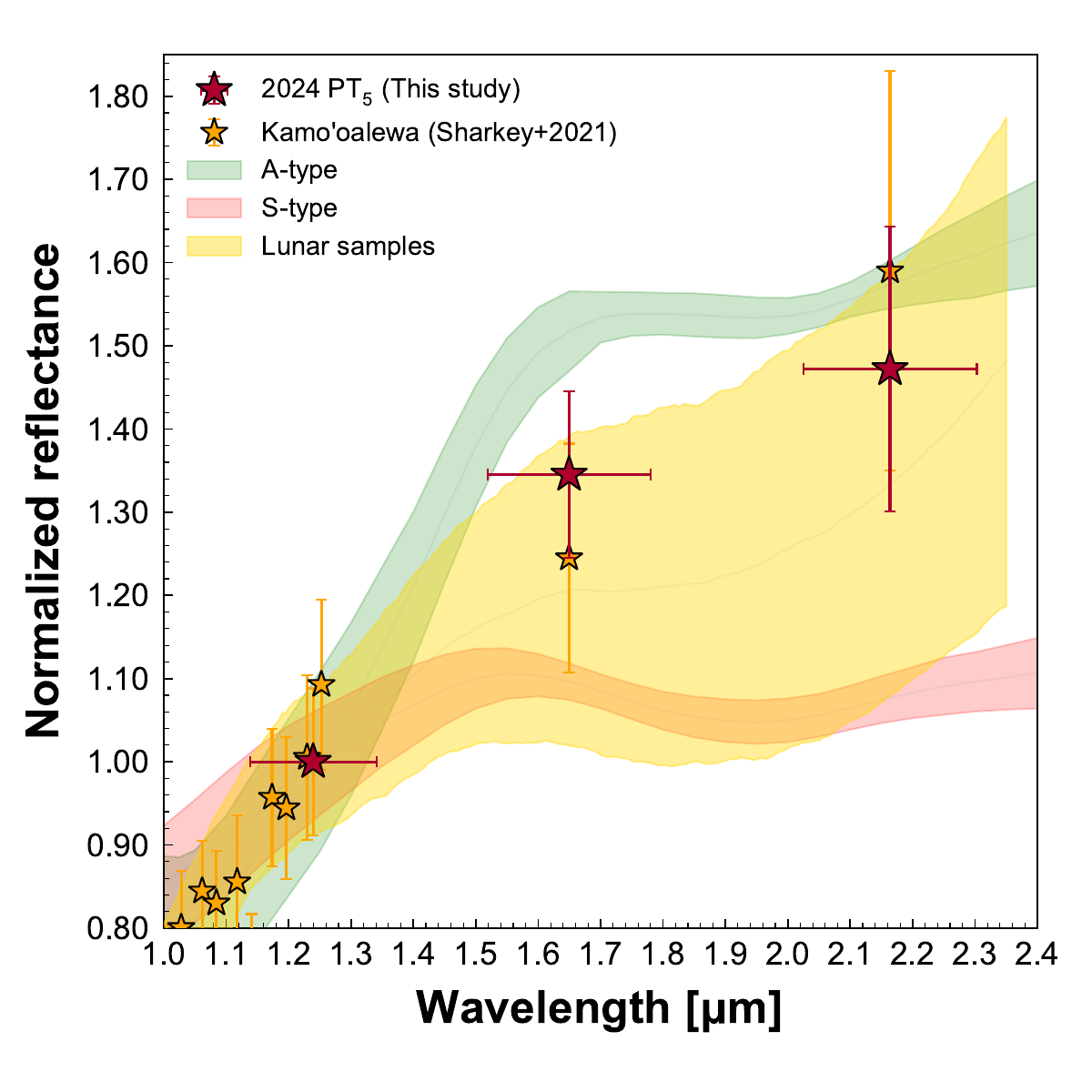}
\end{subfigure}
\caption{
Reflectance spectrum of \PT and \Kamo \citep{Sharkey2021}.
Also, Mahlke templates of A- and S-type asteroids are shown \citep{Mahlke2022}.
Shaded areas indicate the standard deviations of the template spectra. 
The lunar rock core sample colors are also shown \citep{Isaacson2011}.
Shaded areas correspond to the 1$\sigma$ envelope of the compiled spectra.
}
\label{fig:ref}
\end{figure*}

The reflectance spectra of \PT in Fig. \ref{fig:ref} were calculated with 
the derived and solar colors in the same manner as \cite{Beniyama2023b, Beniyama2023c}.
For $g$, $r$, $r$, and $z$ measurements, 
The reflectances at the central wavelength of the $g$, $V$, $r$, $i$, $z$, $J$, $H$, and $Ks$ bands, 
$R_g$, $R_V$, $R_r$ $R_i$, $R_z$, $R_J$, $R_H$ and $R_{K_s}$, were calculated as:
\begin{equation}
R_x = 10^{-0.4[(x-y)_{\mathrm{PT_5}}-(x-y)_\odot]},
\end{equation}
where $x$ and $y$ are the indices of the bands.
The reflectance at $x$-band is normalized relative to that of the $y$-band.
$(x-y)_\mathrm{PT_5}$ is the color of \PT,
and 
$(x-y)_\odot$ is the color of the Sun.
We referred to the absolute magnitude of the Sun in the AB system as 
$g=5.03$, $r=4.64$, $i=4.52$, and $z=4.51$,
and in the Vega system as
$V = 4.81$, $J=3.67$, $H=3.32$, and $K_s=3.27$ \citep{Willmer2018}.

Using the seven band measurements in the visible to near-infrared, we tried to assess our \PT's spectrum.
However, we cannot naturally connect our spectrum's visible and near-infrared portions.
This could be due to the difference in observing geometries between Seimei/TriCCS and Keck/MOSFIRE 
observations as shown in Fig. \ref{fig:loc};
the
cross section of the asteroid might have changed after Seimei/TriCCS observations 
owing to the changing aspect angle \citep{Kwiatkowski1992, Jackson2022, Carry2024}.
Thus, we consider the visible and near-infrared measurements separately. 
We normalized reflectance spectra to unity at the $V$ (0.5511~$\mu$m) and $J$ bands (1.2393~$\mu$m) \citep{Willmer2018} 
in the visible and near-infrared wavelength portions of our spectrum as seen in the left and right panels of Fig.~\ref{fig:ref}.
Our visible spectra derived from colors are consistent 
with results of spectrophotometry in \citet{Bolin2025_PT5} and \citet{Kareta2025}\footnote{Downloaded from the Mission Accessible Near-Earth Objects Survey (MANOS) website, \url{https://manos.lowell.edu/}.}
as well as visible spectra shown in \citet{Kareta2025} and \citet{Marcos2025}.

In Fig. \ref{fig:ref}, 
Mahlke templates of S- and A-type asteroids \citep{Mahlke2022},
a spectrum of \Kamo \citep{Sharkey2021},
and lunar rock core samples \citep[][]{Isaacson2011}
are shown for comparison.
The envelopes of lunar rock core samples are provided by \citet[][]{Isaacson2011} by taking the average of lunar samples.
In visible wavelengths, our spectrum matches well with 
both asteroids, such as S- and A-types templates, and lunar rock core samples.
In the near-infrared wavelengths, the closest match is to lunar rock samples. 
These findings are consistent with previous studies \citep{Bolin2025_PT5, Kareta2025, Marcos2025}. 
When considered separately, our reflectance spectrum is almost identical to that of \Kamo in visible and near-infrared wavelengths. 
However, the combined spectrum is less red than that of \Kamo, as reported in \citet{Kareta2025}.

%
The orbital evolution of \PT before 1937 cannot be reconstructed
due to a close encounter with Earth that causes orbital divergence \citep{Kareta2025, Marcos2025}.
Therefore, to investigate the origin of \PT in terms of a dynamical point of view,
we need to compare the likelihood of lunar ejecta origin and NEO population origin. 
\cite{Fedorets2020b} explored the lunar ejecta origin of the minimoon 2020~CD$_3$ 
by estimating the likelihood based on examining the contemporary production rate of 
small craters on the Moon with plausible assumptions regarding 
their properties (e.g., ejecta/impactor diameter ratio, ejecta speed).
They ruled out a lunar ejecta origin for 2020~CD$_3$, and 
concluded that minimoon capture from the NEO population \citep{Granvik2012, Fedorets2017} 
is a dominant mechanism for maintaining the minimoon steady-state population.

Recently, \cite{Jedicke2025} assessed the likelihood of both NEA origin \citep{Granvik2012, Fedorets2017} 
and lunar ejecta origin for mini-moons using the steady-state size frequency distribution (SFD).
As they discussed in the paper, there is systematic uncertainty of at least a few orders of 
magnitude on their SFD of lunar ejecta, taking the crater scaling relation, 
ejecta SFD, and ejecta size-speed relation into account.
Thus, due to the model's large uncertainties, we cannot conclude whether NEA origin or lunar origin is preferred.

We cannot conclude the origin of \PT with our current knowledge from the discussion above, 
even when spectral and dynamical evidence are considered together.
This conclusion is also valid for the minimoon 2020~CD$_3$.
The visible geometric albedo of the Moon was derived in previous studies \citep[e.g., Table 4 of ][]{Warell2004}.
The estimated bulk visible geometric albedo of the Moon is $\sim0.15$.
The area ratio of lunar mare and highland regions is around 43~\% and 57~\%, and the albedo for mare and highland is $\sim0.1$ and $\sim0.2$ \citep{Korokhin2007}.
If \PT is of lunar origin, \PT could have originated from a bright highland region, as \Kamo could be
from the lunar Giordano Bruno crater \citep{Jiao2024}.
The preference of the highland region is consistent with the argument from band center analysis in \cite{Kareta2025}.

The Rubin Observatory Legacy Survey of Space and Time in Chile \citep[LSST,][]{Ivezic2019}
is expected to observe more than five million asteroids, including 100,000 NEAs
during its ten-year survey \citep{Schwamb2023, Bolin2025TS}.
Since there may be at least one minimoon of 1~m diameter (not of lunar ejecta origin, but from NEO population) at any given time \citep{Granvik2012},
the LSST is expected to discover minimoons as well \citep{Bolin2014, Fedorets2020a}. 
The preparations for follow-up observation of future minimoon objects is thus highly encouraged,
especially in both visible and near-infrared wavelengths are preferable as discussed in \cite{Kareta2025}.

\section{Conclusions} \label{conc}
We conducted visible to near-infrared multicolor photometry of \PT 
with Seimei/TriCCS and Keck/MOSFIRE in January 2025. 
We derive color indices for \PT of 
\gr, \ri, \rz, \YJbyBB, \JHbyBB, and \HKsbyBB,
which indicate that \PT is a typical S-type asteroid.
Our three-days lightcurves are indicative of a tumbling motion of \PT. 
A geometric albedo for \PT of \pVlinear was derived from the slope of its photometric phase curve, which is a fairly good match to S-type and Q-type NEAs. 
The albedo of \PT has not been previously reported.
Using the $H$-$G$ model, we derived an absolute magnitude $H_V$ of \HVHG and a slope parameter $G_V$ of \GV in V-band.
The diameter of \PT is estimated to be \DiamVHG using the albedo and absolute magnitude. 
Based on the physical properties described above,
our results are broadly consistent with a lunar spectral type, with similarity to other co-orbitals with lunar-like spectra 
such as 2020 CD$_3$ and Kamo`oalewa, and in agreement with existing works \citep{Bolin2025_PT5, Kareta2025, Marcos2025}.
This study shows that multiple observations at a wide range of phase angles using a medium-class telescope are essential to further investigate the origin of the minimoon \PT.

\begin{acknowledgements}
We would like to thank Dr.~Alessandro Morbidelli for insightful discussions and helpful comments.
We thank Dr.~Katsuhiro Murata for observing assistance using Seimei telescope.
The authors are grateful to our reviewer Dr.~Marcel Popescu for 
constructive comments on the manuscript.
We acknowledge Dr.~Robert Jedicke and Luis Langermann for the discussion 
regarding the steady state population of Earth's minimoons of lunar provenance
and albedo of the Moon, respectively.
J.B. thanks Dr.~Theodore Kareta and Dr.~Benjamin N. L. Sharkey
for their constructive feedback on the spectra of \PT and \Kamo, respectively.
J.B. is grateful to Dr.~Max Mahlke for the development of, 
and valuable comments on, classy, a tool for exploring, downloading, analyzing, 
and classifying asteroid reflectance spectra.
This work was supported by JSPS KAKENHI Grant Numbers JP23KJ0640 and 25H00665. 
This work was supported by the French government through the France 2030 
investment plan managed by the National Research Agency (ANR), as part of the
Initiative of Excellence Université Côte d’Azur under reference number ANR- 15-IDEX-01.
Some of the data presented herein were obtained at Keck Observatory, which is a private 501(c)3 
non-profit organization operated as a scientific partnership among the California Institute of Technology, the University of California, 
and the National Aeronautics and Space Administration. The Observatory was made possible by the generous financial support of the W. M. Keck Foundation.
Keck Observatory is located on Maunakea, land of the K$\mathrm{\bar{a}}$naka Maoli people, 
and a mountain of considerable cultural, natural, and ecological significance to the indigenous Hawaiian people. 
The authors wish to acknowledge the importance and reverence of Maunakea and express gratitude for the opportunity to conduct observations from the mountain.
This research has made use of LTE's Miriade VO tool.
\end{acknowledgements}

\bibliographystyle{aa} 
\input {main.bbl}

\end{document}

%% file: main.bbl
\newcommand{\noopsort}[1]{} \newcommand{\singleletter}[1]{#1}